# Insights on Numerical Damping Formulations Gained from Calibrating Two-Dimensional Ground Response Analyses at Downhole Array Sites

Nishkarsha Dawadi [a,*], Kami Mohammadi [b], Mohamad M. Hallal [c], Brady R. Cox [a]

[a] Department of Civil and Environmental Engineering, Utah State University, Logan, Utah, USA
[b] Department of Civil and Environmental Engineering, University of Utah, Salt Lake City, Utah, USA
[c] Department of Civil and Environmental Engineering, University of California, Berkeley, California, USA

**ABSTRACT**

Appropriately modeling seismic wave attenuation is critical for ground response analyses (GRAs), which aim to replicate local site effects observed in ground motions. Accurately predicting even small-strain site effects remains challenging because theoretical transfer functions (TTFs) from GRAs often overestimate empirical transfer functions (ETFs) when the small-strain damping ratio ($D_{min}$) is set equal to laboratory-obtained values. Previous studies addressed this by increasing $D_{min}$ in one-dimensional (1D) GRAs to account for apparent damping mechanisms that cannot be inherently modeled in 1D. These attempts have improved predictions of fundamental-mode amplitudes, but often result in overdamping at higher modes. This study investigates more direct modeling of apparent damping using two-dimensional (2D) GRAs at four downhole array sites: Delaney Park (DPDA), I-15 (I15DA), Treasure Island (TIDA), and Garner Valley (GVDA). At each site, three numerical damping formulations were evaluated: Full Rayleigh, Maxwell, and Rayleigh Mass, each implemented with both a conventional $D_{min}$ and an inflated $D_{min}$ ($m \times D_{min}$) obtained from site-specific calibration. Results show that the appropriate $D_{min}$ multiplier ($m$) correlates well with the velocity contrast of a site. When applied with inflated $D_{min}$, Full Rayleigh and Maxwell damping systematically overdamped higher modes, and Maxwell damping also shifted modal peaks, particularly at higher frequencies. In contrast, Rayleigh Mass damping provided the closest match to ETFs at three sites, reducing transfer function misfit by 27-44% relative to Full Rayleigh and 44-49% relative to Maxwell. Rayleigh Mass damping was also computationally efficient, allowing average timesteps more than eight times larger than those with Full Rayleigh. These findings demonstrate that inflated $D_{min}$ can account for unmodeled attenuation in 2D GRAs, particularly at sites with low velocity contrast, and that frequency-dependent formulations such as Rayleigh Mass damping can more effectively capture site response than traditional frequency-independent approaches, underscoring the need to reevaluate the long-standing practice.

* *Corresponding author: nishkarsha.dawadi@usu.edu.* *This manuscript is in press in Earthquake Spectra.*

# 1. Introduction

Ground response analysis (GRA) is a numerical or analytical procedure used to evaluate how seismic waves are modified as they propagate from bedrock, through overlying soil and rock layers, to the ground surface. As the waves travel upward, their amplitude, frequency content, and duration are influenced by the stiffness and attenuation properties of the materials they pass through, as well as by energy dissipation through wave scattering mechanisms like mode conversion and diffraction. These modifications, collectively referred to as site effects, are central to seismic hazard assessment and are critical for designing earthquake-resistant infrastructure. While one-, two-, and three-dimensional (1D, 2D, and 3D) GRAs have been developed to simulate the complexity of the subsurface and seismic wave propagation patterns, the overwhelming majority of GRAs presently performed in practice utilize 1D approaches. These 1D GRAs simplify the subsurface into a stack of horizontal layers of infinite lateral extent, each with uniform thickness and homogeneous properties, and only model vertically propagating, horizontally polarized shear waves. Despite widespread use, 1D GRAs exhibit notable limitations, as recent studies have shown that, on average, recorded ground motions at more than 50% of borehole array sites are poorly modeled using 1D GRAs, with actual percentages ranging from approximately 30% to 80% (e.g., Hallal et al. 2022b; Tao and Rathje, 2020; Afshari and Stewart, 2019; Pilz and Cotton, 2019; Kim and Hashash, 2013; Kaklamanos et al., 2013; Thompson et al., 2012; Thompson et al., 2009). Most of these studies have compared 1D GRA predictions in the form of theoretical transfer functions (TTFs) against small-strain recorded ground motions in the form of empirical transfer functions (ETFs), as the use of small-strain motions isolates wave propagation effects by eliminating the complications introduced by modeling nonlinear soil behavior. A consistent observation across these studies is that 1D GRAs tend to significantly overestimate amplification near a site's fundamental ($f_0$) and higher-mode frequencies. This over-estimation of amplification in 1D GRA has been attributed to improperly modeling wave-scattering mechanisms as a result of over-simplifying subsurface heterogeneities that cause energy loss in the actual field conditions.

Seismic wave attenuation arises from three primary mechanisms: intrinsic, geometric, and apparent damping (Zywicki, 1999). Intrinsic, or material damping, represents energy dissipation within soils and rocks through processes such as interparticle friction and viscous fluid–solid interactions. It is commonly measured in laboratory tests (e.g., resonant column, torsional shear) and often assumed to be frequency independent over the 0.1–10 Hz range relevant for site response (Shibuya et al., 1995; Aki and Richards, 1980), though some studies suggest weak frequency dependence (Menq, 2003; Lai and Özcebe, 2016). Geometric damping refers to amplitude decay due to geometric spreading of wavefronts over progressively larger areas/volumes as waves propagate further from a source. Finally, apparent damping arises from wavefield phenomena such as wave scattering, reflections at interfaces, and mode conversions, induced by subsurface heterogeneity (Rix et al., 2000; Spencer et al., 1977; O'Doherty and Anstey, 1971). This form of damping is strongly site-specific and often dominates in situ observations, where all three mechanisms act simultaneously, complicating the interpretation of small-strain in-situ damping (Abbas et al., 2025).

To better represent energy loss from unmodeled apparent damping, several researchers have proposed increasing the small-strain damping ratio ($D_{min}$) used in 1D GRAs as a practical means to approximate the influence of wave scattering and reduce the overestimation of site response. Different methods have been proposed for adjusting damping, such as increasing $D_{min}$ values to the range of 2%–5% (Cabas et al., 2017; Tsai and Hashash, 2009), adding a fixed increment (e.g., 1%–4%) to laboratory $D_{min}$ values (Yee et al. 2013), applying a multiplier between 2 and 6 to $D_{min}$ values (Pretell et al., 2023a; Tao and Rathje, 2019; Zalachoris and Rathje, 2015), or using alternative damping models based on site-specific parameters like kappa decay (Afshari and Stewart 2019). The study conducted by Tao and Rathje (2019) is notable for its analysis of the required inflation of small-strain damping via a $D_{min}$ multiplier (ɱ) to match observed site response at four downhole array sites. Their results showed that the necessary ɱ value was linked to the local geology, the depositional environment, and the potential for spatial variability in shear wave velocity (Vs), and that ɱ values between 2 and 6 were often necessary to accurately represent recorded small-strain ground motions. Afshari and Stewart (2019) corroborated these findings by investigating the ɱ value necessary to match empirical attenuation estimates derived from the site-specific kappa decay parameter, concluding from a study of 19 downhole array sites in California that the mean and standard deviation of the required ɱ value were 3.6 and 2.4, respectively. While these damping multipliers helped reduce the overestimation of the ETF amplitude at the $f_0$ peak, they often led to over-damping (i.e., underprediction) of amplification at higher-mode frequencies. Hallal et al. (2022b) used the ɱ values suggested by Tao and Rathje (2019) to carry out 1D GRAs at the Treasure Island Downhole Array (TIDA) and Delaney Park Downhole Array (DPDA) sites, and found that while the amplitude of the $f_0$ peak matched closely to the measured peak, higher-mode peaks were significantly underpredicted. In separate research, Kaklamanos et al. (2020) applied a factor of 0.5 to reduce $D_{min}$, aiming to address the underestimation of amplification at higher frequencies rather than to correct the overestimation near the site's $f_0$.

Other methods proposed to implicitly account for spatial variability in a 1D GRA framework, such as randomizing Vs profiles (Toro, 2022, 1995), randomizing cumulative shear wave travel time profile (Hallal et al., 2022a; Passeri et al., 2020) and using multiple Vs profiles from surface wave testing (Teague et al., 2018), have also often led to underprediction of amplification at higher frequencies (Chang et al., 2022; Teague et al., 2018). Although most studies have focused on using either Vs randomization or damping modification independently, a few have also explored the combined effect of both approaches. Rathje et al. (2010) noted an additive effect when multiple sources of site variability were incorporated, suggesting that each method should be applied at lower levels when used together. Rodriguez-Marek et al. (2017) applied both approaches to account for spatial variability and wave scattering, using a lower ɱ value of 2.11 and capping $D_{min}$ at 5%, with Vs randomization based on a modified Toro (1995) model that excluded layer thickness randomization. While this combined approach may improve the handling of spatial variability in 1D GRAs, recent studies (e.g., Pretell et al., 2023b, 2023a) have proposed a more standardized framework for its application. Specifically, Pretell et al. recommend the use of a consistent damping multiplier (ɱ ≈ 3) in combination with Vs randomization ($\sigma_{ln,Vs} \approx 0.25$), calibrated using borehole array data to minimize prediction errors and reduce bias in 1D site response analyses. Although this framework represents a significant step toward standardization, it does not provide a path forward for using different ɱ values based on site-specific subsurface variability. Researchers have used stochastic random fields at downhole array sites to extend 1D Vs profiles into 2D and 3D models, helping to reduce overamplification of the $f_0$ peak by introducing spatial variability that lowers amplitudes and broadens the peak (de la Torre et al., 2022; Hu et al., 2021; Thompson et al., 2009). However, they mainly capture in-layer variability and can miss larger-scale features such as stratigraphic shifts, dipping layers, or abrupt bedrock depth changes.

True site-specific 2D or 3D subsurface models could, theoretically, improve site-response predictions by inherently incorporating all damping mechanisms into multi-dimensional GRAs; however, developing such models over sufficiently large areas and depths remains a significant challenge. Hallal and Cox (2021a) developed a Horizontal-to-Vertical-spectral-ratio (H/V) Geostatistical Approach to create large-scale, site-specific, pseudo-3D subsurface Vs models at the TIDA and the DPDA sites, and used those models to determine the area influencing site response. They found that incorporating large spatial areas improved the alignment between the predicted TTFs and observed ETFs and suggested that a minimum surface area of 400 m × 400 m should be considered for incorporating spatial variability in GRAs. In a subsequent study, Hallal and Cox (2023) explored the lateral extent of the subsurface influencing seismic site response using 2D GRAs at the TIDA site. By analyzing lateral variability down to a depth of 150 m across the TIDA site using cross-sections from a pseudo-3D Vs model, they found that subsurface spatial variability up to 1 km away from the TIDA sensors influenced the recorded site response. This approach allowed them to better model complex phenomena, such as the broadening of the $f_0$ peak and appearance of a secondary amplification peak adjacent to $f_0$, leading to more accurate site response predictions than those from 1D GRAs. Dawadi et al. (2024) conducted similar research at the DPDA site, performing 2D GRAs with varying lateral extents and azimuths. They found that the site response was sensitive to the spatial extent considered in the 2D analyses, with increasing lateral extent causing significant changes in the site response. However, their study at DPDA revealed that modeling spatial variability in 2D GRAs did not fully address the significant overestimation of the recorded ETF amplitude at $f_0$. While pseudo-3D Vs models developed from the H/V geostatistical approach can capture wave scattering off prominent impedance contrasts (e.g., soil-bedrock), leading to broadening of the $f_0$ peak and, at times, to secondary peaks observed in the ETFs (Dawadi et al., 2024; Hallal and Cox, 2023), the pseudo-3D Vs approach relies heavily on a single Vs profile from the downhole array, which may miss lenses or other discontinuous layer boundaries away from the array that would increase apparent damping. Thus, using inflated $D_{min}$ in conjunction with modeling spatial variability inferred by a pseudo-3D Vs model may be required to better match ETF amplitudes when performing 2D/3D GRAs.

In this study, we conduct 2D GRAs at four well-characterized downhole array sites: DPDA in Anchorage, Alaska, I-15 Downhole Array (I15DA) in Salt Lake City, Utah, TIDA in Northern California, and Garner Valley Downhole Array (GVDA) in Southern California. We evaluate three numerical damping formulations: (1) Full Rayleigh Damping, (2) Maxwell Damping, and (3) Rayleigh Mass Damping. For each damping formulation, two small-strain damping values are considered: (1) the conventional $D_{min}$ derived from empirical relationships based on laboratory testing, and (2) an inflated $D_{min}$ ($ɱ \times D_{min}$) obtained through site-specific calibration aimed at improving the match between TTF and ETF amplitudes at the $f_0$ peak. These combinations of numerical damping formulations and small-strain damping values are assessed for their ability to reproduce observed site response at each downhole array. Analyses are conducted in FLAC3D version 9.0 (Itasca Consulting Group 2023), and the computational efficiency of each formulation is also compared. The objective is to determine the optimal numerical damping formulation and $ɱ$ value at each downhole array, improving the accuracy of site response predictions. We also examine the efficacy of widely used Full Rayleigh damping, the reliability of virtually frequency-independent Maxwell damping, and the practicality of frequency-dependent Rayleigh Mass damping in capturing real-world site behavior.

## 2. Description of the Downhole Array Sites

The four downhole array sites used in this study, DPDA, I15DA , TIDA, and GVDA, were selected to represent a diverse range of geologic settings and seismic site responses. The descriptions provided below for each site contain a brief overview of the subsurface geology, downhole array configuration, and invasive and non-invasive dynamic site characterization data available. Additional details on non-invasive site characterization data are available for DPDA, TIDA, and GVDA in Hallal et al. (2025), and for I15DA in Cox et al. (2025). To facilitate presentation of information in the sections below, Figure 1 provides the 1D soil stratigraphy and a 1D Vs profile for each downhole array. These 1D Vs profiles were used in conjunction with the H/V geostatistical approach (Hallal and Cox, 2021a) to develop a large-scale pseudo-3D Vs model at each site to allow multi-dimensional GRAs (refer to Figure 2). A brief description of the H/V Geostatistical Approach is provided herein, and the reader is referred to Hallal and Cox (2021a) for additional details.

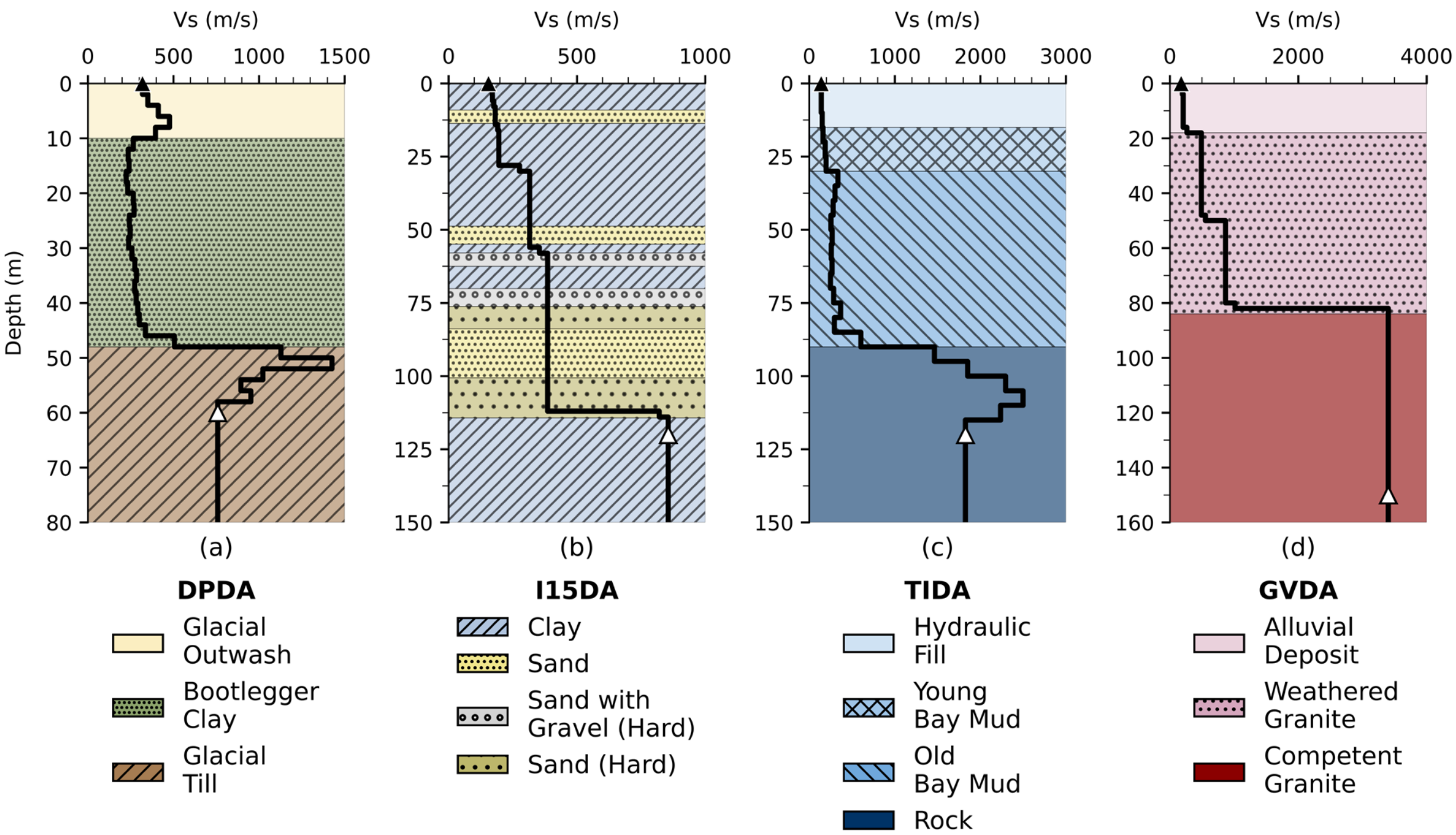

**Figure 1.** 1D Vs profiles used in this study to develop the pseudo-3D Vs models at: (a) DPDA, (b) I15DA, (c) TIDA, and (d) GVDA, presented along with the corresponding 1D soil stratigraphy. Solid and hollow triangular symbols denote the surface and deepest downhole sensors, respectively, at each site.

The H/V geostatistical approach requires only a single measured Vs profile and a large number of simple, cost-effective, H/V noise measurements to develop a pseudo-3D Vs model that can be used to incorporate realistic Vs spatial variability into site response studies. It relies on geostatistical interpolation techniques to develop a uniform-gridded map of $f_0$ values obtained from spatially-distributed H/V measurements (i.e., $f_{0,H/V}$) to show how resonant frequencies vary across the site. These $f_{0,H/V}$ values at each grid point are then used to develop scaling factors that are applied to a 1D Vs profile measured at the downhole array location as a means to develop a modified/scaled 1D Vs profile at each grid point in the map. When these 1D Vs profiles are placed in their respective spatial positions, they form a pseudo-3D Vs model that can extend to kilometers laterally in each direction around the array. The scaled 1D Vs profiles at each grid location in the pseudo-3D Vs model have an $f_{0,H/V}$ value that is consistent with the changes in resonant frequency around the site, which is a key feature for modeling seismic site response. Pseudo-3D Vs models have previously been developed at DPDA and TIDA (Hallal and Cox, 2021a) and have been shown to be effective at incorporating spatial variability into site response studies (Dawadi et al., 2024; Hallal and Cox, 2023; Hallal et al., 2022b; Hallal and Cox, 2021b). The DPDA model was subsequently refined by Dawadi et al. (2024) using additional H/V data. In the present study, these same approaches have been used to develop pseudo-3D Vs models for I15DA and GVDA. Figure 2 presents the pseudo-3D Vs models for all four sites. Additional information for each downhole array site is provided in the following site-specific sections.

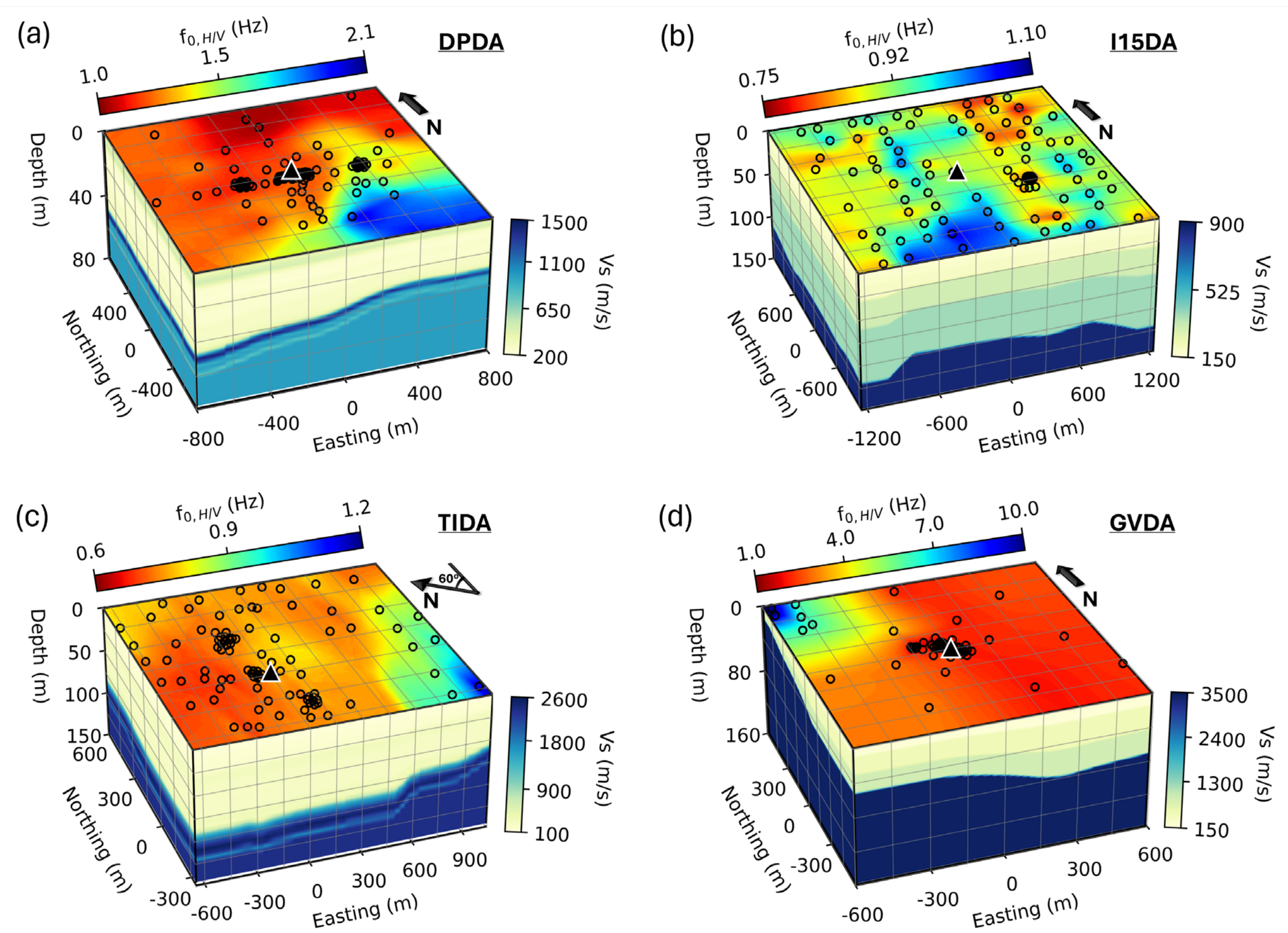

**Figure 2.** Pseudo-3D Vs models developed using the H/V geostatistical approach at: (a) DPDA, (b) I15DA, (c) TIDA, and (d) GVDA. H/V measurement locations at each site are indicated by black circular symbols, and the locations of the downhole array surface sensor at each site are indicated by a black solid triangular symbol.

### *2.1 Delaney Park Downhole Array (DPDA)*

The DPDA site is located in downtown Anchorage, Alaska. The subsurface profile consists of approximately 10 to 15 m of glacial outwash, underlain by about 30 m of Bootlegger cove clay, and followed by glacial till (Combellick 1999). The 1D soil stratigraphy at the downhole array is shown in Figure 1a. The Bootlegger Cove Clay is a Pleistocene-age deposit of interbedded silty clays, clayey silts, and fine sands, containing both cohesive and non-cohesive facies. Cohesive zones are highly sensitive and prone to strength loss, whereas non-cohesive facies are more susceptible to liquefaction (Badal et al. 2004).

Strong-motion sensors are installed at depths of 0, 4.6, 10.7, 18.3, 30.5, 45.4, and 61 m; however, only the surface (0 m) and deepest (61 m) sensors, represented by solid and hollow black triangular symbols in Figure 1a, respectively, were used in this study to evaluate site response. A Vs profile developed from seismic downhole measurements by Thornley et al. (2019), illustrated using a black solid line in Figure 1a, was used to construct the pseudo-3D Vs model. The pseudo-3D Vs model at DPDA (Figure 2a) spans 1.6 × 1.6 km laterally and extends to a depth of 80 m. A total of 108 H/V spectral ratio measurements, indicated by black circular symbols in Figure 2a, were used to develop the model. The pseudo-3D Vs model at DPDA contains 25.6 million, 2-m cubic elements with velocities ranging from approximately 200 m/s near the surface to 1450 m/s at the interface between the Bootlegger cove clay and glacial till. From Figure 2a, it can be seen that higher velocities associated with glacial till occur closer to the surface in the southeastern portion of the model, consistent with the stratigraphy reported by Combellick (1999).

### *2.2 I-15 Downhole Array (I15DA)*

The I15DA site is located near the intersection of I-15 and I-80 in South Salt Lake City, Utah. The subsurface consists of more than 100 m of fine-grained lacustrine deposits from ancient Lake Bonneville, including interbedded clay, silts, sands, and occasional gravel lenses, as shown in Figure 1b. These lakebed sediments are representative of conditions across much of northern Utah's valleys.

Strong-motion sensors are installed at depths of 0, 7.6, 18.3, 48.8, and 119.8 m; however, only the surface (0 m) and deepest (119.8 m) sensors were used in this study to evaluate site response. A Vs profile developed by Jackson (2024) from deep surface wave testing using an inversion layering ratio parameterization of 2.0 was used along with 98 H/V spectral ratio measurements to construct the pseudo-3D Vs model for I15DA. The pseudo-3D Vs model at I15DA spans 2.5 × 2.5 km laterally and extends to a depth of 160 m. It contains approximately 125 million, 2-m cubical elements with velocities ranging from approximately 150 m/s at the surface to 900 m/s at the base. As shown in Figure 2b, the subsurface at I15DA is composed of an undulating stiff clay layer that does not dip uniformly in one direction, but varies across the site.

### *2.3 Treasure Island Downhole Array (TIDA)*

The TIDA site is located on a man-made island in San Francisco Bay that was constructed using hydraulic fill placed over natural bay sediments. The surface deposits consist of approximately 15 m of loose, saturated, hydraulic fill composed primarily of fine sand with variable amounts of clay and gravel. This is underlain by approximately 15 m of San Francisco Young Bay Mud, followed by approximately 60 m of Old Bay Mud, which comprises silty sand, gravel, and clay-rich Quaternary deposits. Bedrock is encountered at a depth of approximately 90 m and consists of interbedded sandstone, siltstone, and shale of the Franciscan Formation (Gibbs et al., 1994; Pass, 1994). The 1D soil layering at the site is shown in Figure 1c.

Strong-motion sensors are installed at depths of 0, 7, 16, 31, 44, 104, and 122 m, though only the surface (0 m) and deepest (122 m) sensors were used in this study. A Vs profile developed by Graizer and Shakal (2004) from PS suspension logging was used to construct the pseudo-3D Vs model. The pseudo-3D Vs model at TIDA (Figure 2c) spans 1.7 × 0.9 km laterally and extends to a depth of 150 m. It contains approximately 14.7 million, 2.5-m cubical elements, with Vs values ranging from approximately 140 m/s near the surface to 2600 m/s in the bedrock. A total of 98 H/V spectral ratio measurements were incorporated to develop the pseudo-3D Vs model at TIDA. Unlike the other sites in this study, the TIDA pseudo-3D Vs model is not aligned with geographic north, but oriented approximately 60° west of north. It can be observed in Figure 2c that the pseudo-3D Vs model at TIDA reflects the up-dipping bedrock outcrop of Yerba Buena Island toward the south, in agreement with observed site conditions.

### *2.4 Garner Valley Downhole Array (GVDA)*

The GVDA is located in a narrow valley in Southern California. The upper 18–25 m consists of soft alluvial deposits, including silty sand, clayey sand, and silty gravel deposited in an ancestral lakebed. Beneath this, the profile transitions into a thick sequence of decomposed and weathered granite extending to a depth of approximately 87 m. This is underlain by competent granodiorite bedrock of the Peninsular Ranges Batholith, with a pronounced impedance contrast at the weathered-to-competent bedrock interface (Bonilla et al. 2002). The stratigraphy at the GVDA is shown in Figure 1d.

Strong-motion sensors are installed at depths of 0, 15, 22, 50, and 150 m; however, only the surface (0 m) and deepest (150 m) sensors were used in this study. A Vs profile developed by Teague et al. (2018) from deep surface wave testing using an inversion layering ratio parameter of 3.0 was used to construct the pseudo-3D Vs model. The pseudo-3D Vs model at GVDA (Figure 2d) spans 1.2 × 1.2 km laterally and extends to a depth of 160 m. It contains a total of 28.8 million, 2-m cubical elements, with Vs values ranging from about 200 m/s in the surface layers to over 3400 m/s in the bedrock. A total of 63 H/V spectral ratio measurements were used to develop the pseudo-3D Vs model at GVDA. The northeastern portion of the model indicates shallow bedrock outcrop, consistent with the site conditions, and this feature is expected to strongly influence the site response at GVDA.

## 3. Extraction of Cross-sections

For this study, 2D GRAs were employed to capture subsurface complexities beyond what can be represented in 1D analyses. In particular, 2D cross-sections were extracted to include the strongest lateral variability in stratigraphy, such as dipping bedrock and irregular layer interfaces, which can produce wave scattering effects and strongly influence site response. This approach provides a more realistic representation of local geology while remaining computationally efficient. Although full 3D GRAs could additionally capture azimuthal variability and input incoherence, their substantially higher computational demands made them impractical within the scope of this work. Specifically, testing six damping formulations across four downhole array sites would have required a total of 24 large 3D simulation cases, with runtimes ranging from 24 hours to nearly a week per case when running on the workstation (AMD Ryzen Threadripper PRO 5995WX 64-Cores with a clock-speed of 2.70 GHz) used for the present study, depending on the damping formulation and model size. As a practical alternative, 2D GRAs were employed, and a robust procedure was used to identify the most heterogeneous 2D cross-section at each site to reasonably capture key 3D subsurface features.

Multiple cross-sections were extracted from each pseudo-3D Vs model at 15° azimuthal (Az) intervals measured relative to north. As an example, Figure 3a shows the pseudo-3D Vs model at DPDA with section lines indicating the planes along which cross-sections were extracted. To quantitatively assess heterogeneity, coefficients of variation (CV) were computed for each cross-section. Simply computing a single CV value for the entire cross-section would have been misleading, as it would only capture Vs contrasts without reflecting the effects of dipping or undulating interfaces. Instead, row-wise CVs were first calculated at each element-depth increment in the cross-section, producing a CV profile with depth. Figure 3b illustrates the variation of row-wise CV with depth for different cross-sections extracted from the DPDA pseudo-3D Vs model. The average of each individual row-wise CV profile was then calculated, and the section with the highest average value was identified as the most heterogeneous. Finally, a visual inspection was performed to qualitatively verify that this 2D cross-section captured the most heterogeneous features of the pseudo-3D Vs model.

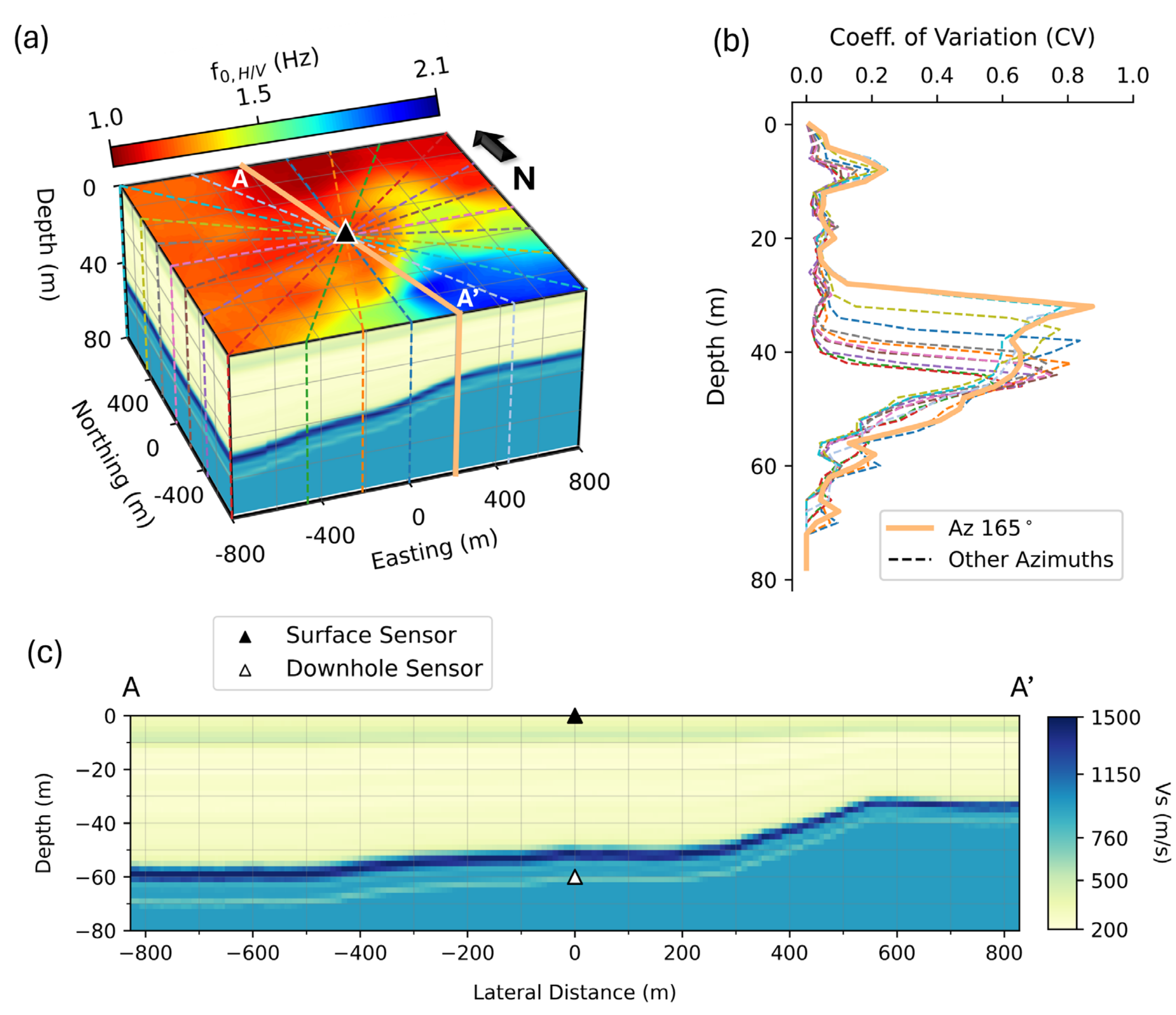

**Figure 3.** Extraction of the most heterogeneous 2D cross-section from the DPDA pseudo-3D Vs model. Shown are: (a) pseudo-3D Vs model with dotted lines indicating planes of different azimuthal cross-sections, (b) coefficient of variation for each azimuthal cross-section, and (c) most heterogeneous azimuthal cross-section (Az = 165°). Solid and hollow triangular symbols denote the surface and downhole sensors, respectively. Section AA' represents the plane of the most heterogeneous cross-section.

At DPDA, the cross-section oriented at 165° azimuth (Az-165°) was identified as the most heterogeneous (Figure 3c). Solid and hollow triangular symbols in Figure 3c denote the surface and downhole sensors, respectively. The center of each cross-section passes through the borehole array, with lateral distance measured relative to this point. As illustrated in Figure 3c, the most heterogeneous cross-section captures the dipping bedrock between +250 m and +550 m lateral distance, which corresponds to the steepest-dipping glacial till feature in the southeastern portion of the DPDA pseudo-3D Vs model (Figure 3a). This dipping till layer is expected to exert the greatest influence on site response at DPDA.

Using the same approach, cross-sections oriented at azimuths of 30°, 165°, and 135° were identified as the most heterogeneous at I15DA, TIDA and GVDA, respectively. Following the recommendations of Hallal and Cox (2023), the full lateral extent of each cross-section, on both sides of the array, was incorporated in all 2D GRAs conducted in this study. Geometrical details on the most heterogeneous cross-sections at each site are provided in the Description of Numerical Model section.

## 4. Recorded Ground Motions and Empirical Transfer Functions

ETFs between the deepest and surface sensors are often calculated as a means to represent the observed site response at a downhole array site. Like previous multi-dimensional GRA studies at downhole array sites (e.g., Hallal and Cox, 2023), the present study focuses only on small-strain GRAs in order to eliminate the need to model additional complexities induced by nonlinear soil behavior and isolate the effects of wave scattering resulting from subsurface spatial variability. Hence, ETFs were derived only from small-strain ground motions recorded at each array and were calculated as the ratio of the Fourier amplitude spectrum (FAS) of the recorded horizontal accelerations at the surface to those at the deepest downhole sensor. To reduce noise and enhance interpretability, all FAS were smoothed using the Konno and Ohmachi (1998) method with a bandwidth coefficient (b-value) of 75 (i.e., minimal smoothing) prior to calculating ETFs. Details regarding the selection of ground motions at each site are provided below.

A summary of the small-strain ground motions used to calculate the ETFs in this study is provided in Table 1. At the DPDA site, 38 low-amplitude ground motions were used to compute the ETFs. This included both horizontal components from 19 unique events, with peak ground accelerations (PGAs) ranging from 0.001 to 0.01g. The corresponding local magnitudes ($M_L$) were between 3.0 and 5.1, and source-to-site distances ranged from 10 to 96 km. These recordings were used to compute the empirical site response between the deepest and surface accelerometers. Additional information on ground motion selection at DPDA is available in Hallal and Cox (2021a). For the I15DA site, a total of 12 horizontal ground motions from six low-amplitude aftershock events were used, with peak ground accelerations (PGAs) ranging from 0.002 to 0.012 g and local magnitudes (ML) between 2.88 and 3.41. These recordings were used to derive the ETFs between the deepest and surface accelerometers. Additional details regarding ground motion selection and processing at I15DA are provided in Jackson and Cox (2026). For the TIDA site, a total of 38 low-intensity events (both horizontal components from 19 unique events) were selected for calculating ETFs. These events had PGAs below 0.05g, allowing the study to isolate linear soil behavior and avoid complexities introduced by nonlinear effects. Further details regarding the ground motion selection and processing at TIDA can be found in Hallal and Cox (2023). At the GVDA site, ground motions for calculating ETFs were originally compiled by Tao and Rathje (2019) and also used by Teague et al. (2018). The present study adopts the same set of 42 events (i.e., 84 horizontal motions) used by Teague et al. (2018). These motions had PGAs ranging from 0.001 to 0.01g, local magnitudes between 3.0 and 5.1, and source distances from 6.6 to 133 km. Readers are referred to Tao and Rathje (2019) for a detailed description of the ground motion selection and to Teague et al. (2018) for further discussion of ETFs at GVDA.

**Table 1.** Summary of small-strain ground motions used to calculate the ETFs in this study.

| Site | Number of Ground Motions | PGA Range (g) | Source |
|---|---|---|---|
| DPDA | 38 | 0.001 – 0.01 | Hallal and Cox (2021a) |
| I15DA | 12 | 0.002 – 0.012 | Jackson and Cox (2026) |
| TIDA | 38 | < 0.05 | Hallal and Cox (2023) |
| GVDA | 84 | 0.001 – 0.01 | Tao and Rathje (2019) |

The individual ETFs calculated from each ground motion were statistically combined into a lognormal median ($LM_{ETF}$), computed as the exponential of the arithmetic mean of the natural logarithms of individual ETFs, which is equivalent to the geometric mean in linear space. To quantify variability, the natural logarithmic standard deviation ($\sigma_{ln,ETF}$) was calculated at each frequency. Because two horizontal ground motion components (east–west and north–south) are available for each event, different approaches can be used to combine the horizontal components prior to computing a single, representative $LM_{ETF}$. In this study, two distinct approaches have been used to combine the horizontal components. First, a direction-independent, averaged representation of site response was computed by taking the lognormal median across the individual ETFs from both horizontal components. This is referred to herein as the combined ETF, denoted as $LM_{ETF,EW\&NS}$, with an associated standard deviation $\sigma_{ln,ETF,EW\&NS}$. This is the approach used to calculate a single representative ETF in the source publications listed in Table 1. Second, an azimuth-specific representation of site response was obtained by rotating (i.e., resolving) the horizontal ground motions along a specific azimuth prior to calculating the individual ETFs. The statistical representation of these azimuth-specific ETFs is referred to herein as the rotated ETF, denoted as $LM_{ETF,Rotated}$, with corresponding standard deviation $\sigma_{ln,ETF,Rotated}$. These rotated ETFs are essential for comparison with 2D GRAs, which inherently represent wave propagation along a specific cross-sectional orientation. The rotated ETFs in this study were resolved along the azimuth corresponding to the most heterogeneous cross-section at each site, as discussed in Section 3. Both representations are used herein for different purposes, as discussed in subsequent sections.

The $LM_{ETF,EW\&NS} \pm 1\,\sigma_{ln,ETF,EW\&NS}$ and the $LM_{ETF,Rotated} \pm 1\,\sigma_{ln,ETF,Rotated}$ bounds for each downhole array site are shown in Figure 4. The modal peak frequencies at all four sites align well between $LM_{ETF,EW\&NS}$ and $LM_{ETF,Rotated}$. The close agreement between the two representations at DPDA, TIDA, and GVDA indicates that directional effects in the recorded ground motions at these sites are minimal, with only subtle differences observed in amplitude and overall spectral shape. However, at I15DA, the variability is more pronounced, with a noticeable increase in $\sigma_{ln,ETF,Rotated}$, leading to wider bounds at the second and third higher-mode peaks. Despite this, the median ETFs remain in good agreement, except within the trough region between the $f_0$ and first higher modes. The differences observed at I15DA may reflect the relatively limited number of ground motions available, leading to increased directional bias, and/or stronger directional effects compared to the other sites. Nevertheless, the frequencies and amplitudes associated with the $f_0$ and first higher mode remain largely consistent. The $f_0$, first higher mode ($f_1$), second higher mode ($f_2$), and third higher mode ($f_3$) identified from $LM_{ETF,EW\&NS}$, along with the maximum resolvable frequencies ($f_s$) of the 3D model developed for each site, are also shown in Figure 4. The maximum resolvable frequency was calculated based on the element size of each pseudo-3D Vs model. According to Kuhlemeyer and Lysmer (1973), the maximum element size in 2D or 3D layered media analyses should be limited to one-eighth of the wavelength of the slowest body wave propagating through the material. Therefore, element sizes were selected to ensure that the soft, near-surface soil layers at each site could be adequately modeled up to a frequency corresponding to the half-amplitude point following the $f_3$ peak in the combined ETF, with a minimum of 10 elements per shortest propagating wavelength. This further refinement of ≥ 10 elements per wavelength was selected to ensure that scattered waves propagating at oblique angles are captured properly. The element size, minimum velocity in each pseudo-3D Vs model, the four frequency peaks, and $f_s$ are summarized in Table 2.

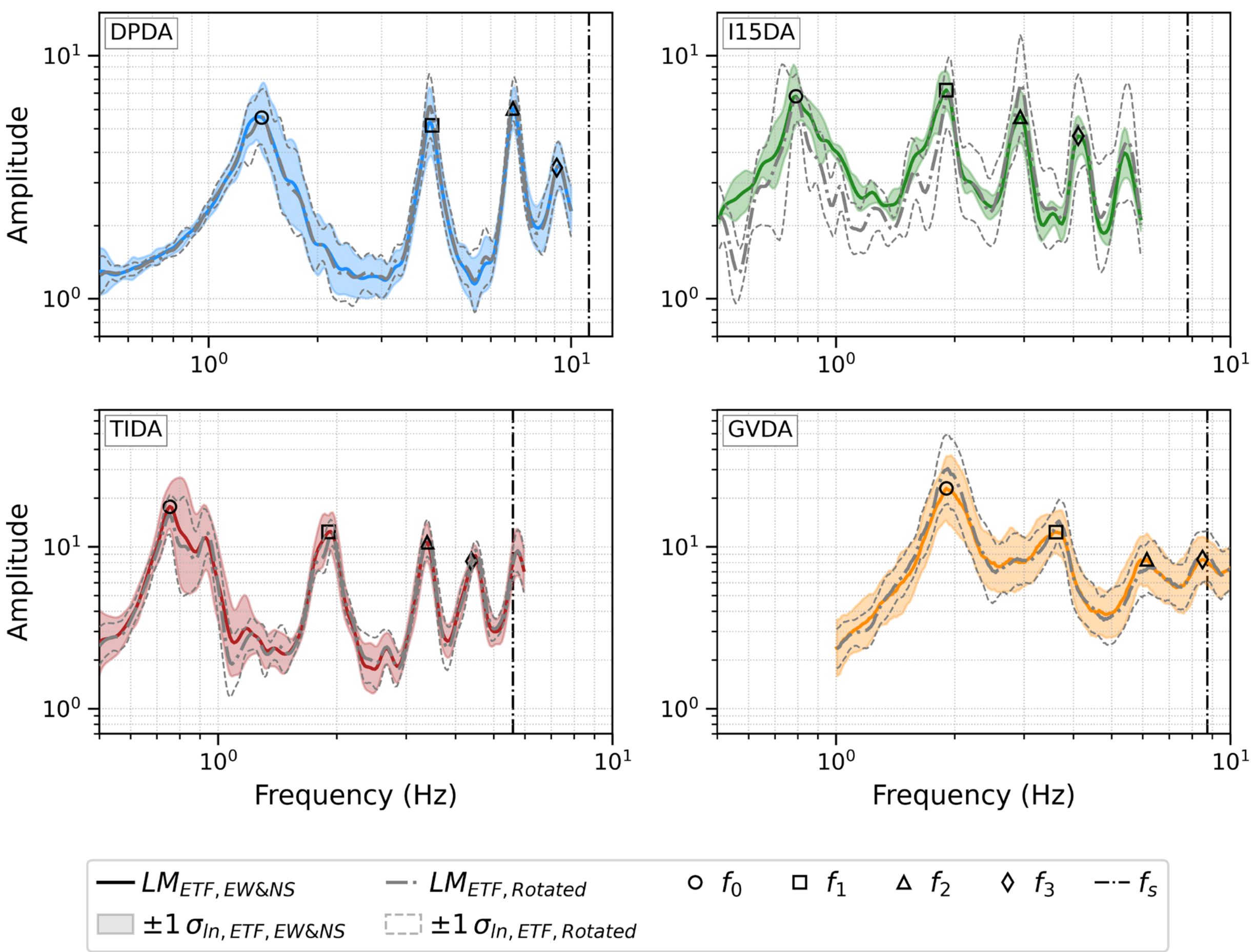

**Figure 4.** Combined ($LM_{ETF,EW\&NS} \pm 1\ \sigma_{ln,ETF,EW\&NS}$) and rotated ($LM_{ETF,Rotated} \pm 1\ \sigma_{ln,ETF,Rotated}$) ETFs at: (a) DPDA, (b) I15DA, (c)TIDA, and (d) GVDA. Also shown are the fundamental mode ($f_0$), first higher mode ($f_1$), second higher mode ($f_2$), third higher mode ($f_3$), and the maximum resolvable frequency ($f_s$).

**Table 2.** Summary of modal frequency peaks and maximum resolvable frequency ($f_s$) at each site

| Site | Element Size (m) | Minimum Vs (m/s) | $f_0$ (Hz) | $f_1$ (Hz) | $f_2$ (Hz) | $f_3$ (Hz) | $f_s$ (Hz) |
|---|---|---|---|---|---|---|---|
| DPDA | 2 | 224 | 1.40 | 4.13 | 6.91 | 9.14 | 11.21 |
| I15DA | 2 | 155 | 0.79 | 1.90 | 2.93 | 4.12 | 7.80 |
| TIDA | 2.5 | 140 | 0.75 | 1.90 | 3.39 | 4.39 | 5.60 |
| GVDA | 2 | 175 | 1.90 | 3.61 | 6.15 | 8.50 | 8.75 |

## 5. Description of Numerical Model

This study employed 2D linear-viscoelastic analyses in FLAC3D version 9.0 (Itasca Consulting Group 2023), a finite-volume software. As noted above, conducting full 3D analyses for all 24 simulation cases was impractical. Instead, 2D GRAs were used to identify the most effective damping formulation at each site, thereby streamlining future 3D GRA efforts by eliminating the need to test multiple formulations. By focusing on linear-viscoelastic modeling, we aim to isolate the effects of spatial variability in subsurface conditions while excluding the complexities introduced by nonlinear soil behavior. This practice of focusing first on accurately modeling linear-viscoelastic site response is consistent with many previous studies at downhole array sites (e.g., Hallal and Cox, 2023; Tao and Rathje, 2019). The validity of the linear-viscoelastic modeling is evaluated through comparison of simulated results with the low-intensity, small-strain ETFs discussed above (refer to Figure 4).

### *5.1 Geometry*

Table 3 provides a summary of the geometric characteristics of the most heterogeneous Vs cross-sections from each site analyzed in this study. As discussed earlier, the element sizes for each site were chosen to ensure adequate resolution of the near-surface soft soil layers up to the frequency corresponding to the half-amplitude point following the $f_3$ peak in the $LM_{ETF,EW\&NS}$, while maintaining at least 10 elements per shortest propagating wavelength (refer to Figure 4). Since the 2D GRAs were conducted using FLAC3D, each cross-section was modeled as a 3D slice with a thickness equal to one element in the out-of-plane direction. To minimize artificial wave reflections not perfectly absorbed by the free-field model boundaries (discussed below), the cross-sections were extended laterally on both sides. The expansion length was selected via a sensitivity study, such that successive domain expansions produced negligible differences in the normalized surface acceleration time histories, indicating that boundary reflections were minimized. Additionally, the models were extended below the deepest downhole sensor to ensure consistency in the boundary conditions between the computed TTFs and ETFs, which account for both upgoing and downgoing wave components, as discussed in a later section. Table 3 includes the original length, extended length, depth, element size, and number of elements for each cross-section. It should be noted that FLAC3D is a finite-volume software in which the discretized cells are formally referred to as “zones”. However, to remain consistent with common practice in the numerical modeling literature, these are referred to as “elements” in this study.

**Table 3.** Summary of geometric characteristics of the 2D cross-sections used in this study

| Site | Original Length (m) | Extended Length (m) | Depth (m) | Element Size (m) | No. of Elements |
|---|---|---|---|---|---|
| DPDA | 1,656m | 2,000m | 80m | 2m | 40,000 |
| I15DA | 2,884m | 4,000m | 150m | 2m | 150,000 |
| TIDA | 1,755m | 3,000m | 150m | 2.5m | 72,000 |
| GVDA | 1,696m | 4,000m | 160m | 2m | 160,000 |

### *5.2 Damping of Soil*

Although the primary focus of this study is on soil damping, other fundamental material properties such as Vs, mass density, and Poisson’s ratio are also required to establish the baseline conditions for the GRAs. Vs values were directly assigned to the FLAC3D model from the discretized cross-sections. A constant mass density of 2,000 kg/m³ was assumed for all layers. Poisson’s ratio was specified as 0.30 for soil layers above the groundwater table, 0.48 for soil layers below the water table, and 0.30 for rock layers with Vs greater than 760 m/s. The groundwater table was assigned at depths of 20 m at DPDA, 20 m at I15DA, 5 m at TIDA, and 2 m at GVDA based on available literatures (Bonilla et al., 2002; Foerster & Modaressi, 2007; Thornley et al., 2019; Youd & Briggs, 2003).

As previously noted, seismic wave attenuation arises from three primary mechanisms: intrinsic, geometric, and apparent damping (Abbas et al., 2025; Zywicki, 1999). In practice, in-situ recordings inherently capture the combined effects of all three mechanisms, complicating efforts to separate intrinsic damping from apparent damping. This makes it challenging to define a single "correct" damping value for use in numerical site response models. Because this section focuses on material properties, the discussion here is limited to intrinsic damping. Geometric and apparent damping, while not directly prescribed as material properties, are discussed later in the context of damping multipliers and their role in reproducing observed site response.

Hardin and Drnevich (1972) investigated the impact of excitation frequency on the maximum shear modulus ($G_{max}$) and $D_{min}$. They found that excitation frequency has a minimal effect on small-strain shear modulus (and thus Vs) across all soil types, but a more noticeable effect on the material damping of clayey soils. Stokoe and Santamarina (2000) corroborated these findings, showing that as excitation frequency increases from 1 Hz to approximately 100 Hz, $G_{max}$ increases by 5-30%, with a more pronounced effect in soils with higher plasticity index (PI) (refer to Figure 5). In contrast, $D_{min}$ exhibits a much stronger frequency dependence, whereby increases in excitation frequency over the same range can result in $D_{min}$ increases of approximately 50–150%, depending on PI. This stronger dependence of damping on frequency, especially in soils with higher fines content (Stokoe et al. 1999), highlights the importance of considering frequency-dependent variations in $D_{min}$ when selecting appropriate values for modeling based on laboratory testing.

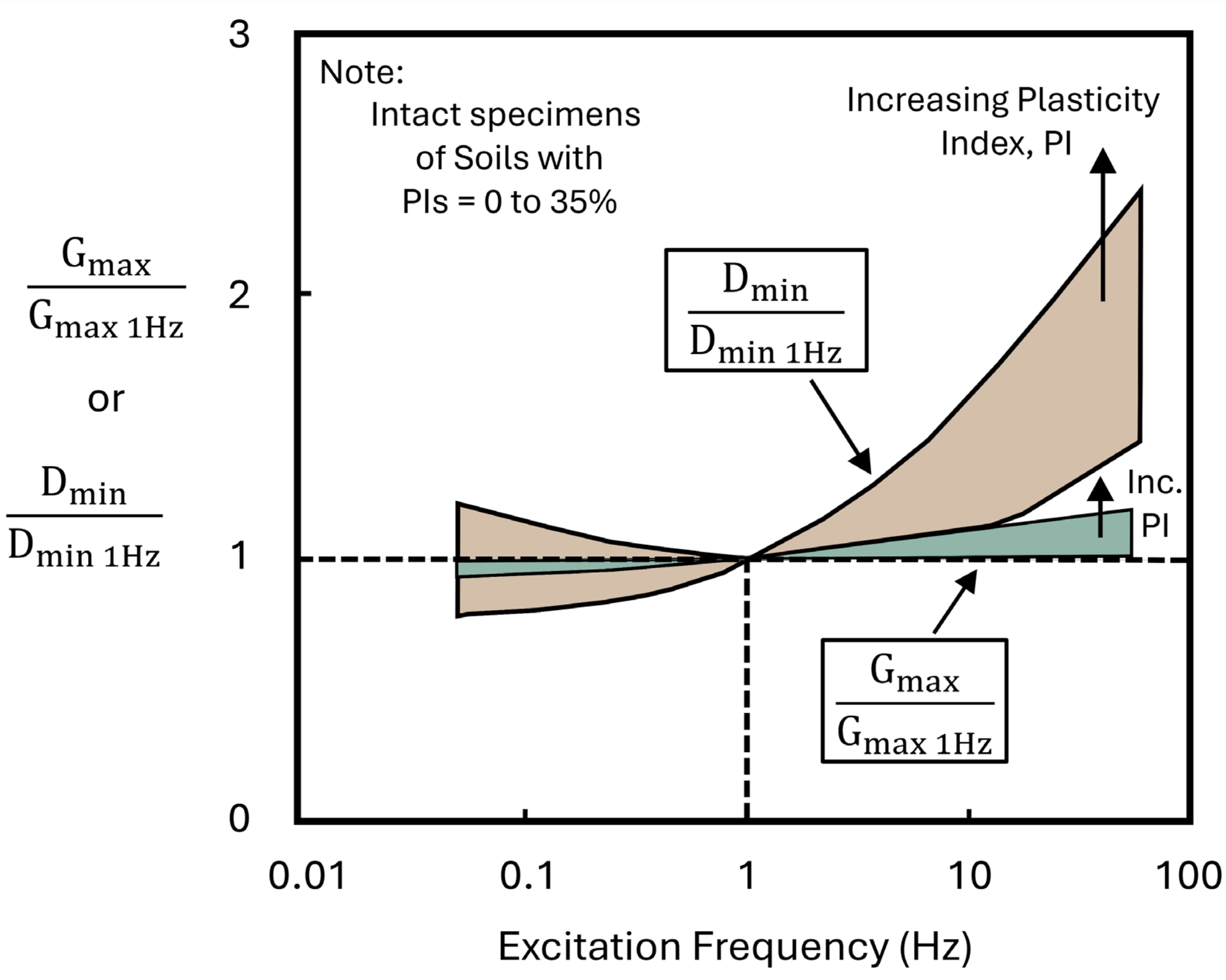

**Figure 5.** General Effect of Excitation Frequency on Small-Strain Shear Modulus, $G_{max}$, and Small-Strain Material Damping Ratio in Shear, $D_{min}$ (after Stokoe and Santamarina, 2000).

The empirical equations proposed by Darendeli (2001), one of the most widely used methods for estimating $D_{min}$, also account for the influence of excitation frequency. The method was developed from an extensive laboratory database consisting of undisturbed soil specimens obtained from geotechnical sites across Northern California, Southern California, South Carolina, and Lotung, Taiwan. The database included sands, silts, and clays tested over a wide range of depths and under varying isotropic confining pressures, providing a comprehensive set of conditions for developing the model. All tests were performed at the University of Texas at Austin using resonant column and torsional shear devices. The underlying dataset spans a broad range of loading frequencies, from low frequencies associated with torsional shear testing to higher frequencies typical of resonant column testing (on the order of ~0.05 to 80 Hz), thereby capturing the frequency-dependent behavior of small-strain damping. Based on this dataset, Darendeli (2001) formulated an empirical procedure that expresses $D_{min}$ as a function of plasticity index, mean effective confining pressure, stress history, and excitation frequency. The proposed equation for estimating $D_{min}$ is:

$$D_{min} = (0.8005 + 0.0129 \times PI \times OCR^{-0.1069}) \times {\sigma_0}'^{-0.2889}[1 + 0.2919\, ln(f_{exc})] \quad (1)$$

where PI is the plasticity index, OCR is the overconsolidation ratio, ${\sigma_0}'$ is the mean effective stress (in atm), defined as the average of the principal effective stresses, and $f_{exc}$ is the excitation frequency in Hz. In most practices, $f_{exc}$ of 1 Hz is assumed when calculating $D_{min}$. However, when a different $f_{exc}$ is used, the calculated $D_{min}$ varies. For example, with a PI of 30, an OCR of 1.5, and a ${\sigma_0}'$ of 20 kPa, the $D_{min}$ values are calculated as 1.87% for an $f_{exc}$ of 1 Hz and 2.47% for that of 3 Hz. Although this difference may appear minor, it can significantly affect results when combined with damping multipliers meant to account for apparent damping in lower-dimensional GRAs.

The ETFs for the four sites in this study (refer to Figure 4) suggest that seismic energy is more evenly distributed across logarithmic frequency bands than across arithmetic intervals. Accordingly, $D_{min}$ in this study was calculated using an excitation frequency of $f_{exc}$ = 3 Hz , which represents the approximate logarithmic median of the frequency distribution of the first four modes of the ETFs across the four sites. All references to $D_{min}$ throughout this paper correspond to values evaluated at $f_{exc}$ = 3 Hz. For comparison, depth-dependent $D_{min}$ profiles were also computed using a conventional excitation frequency of 1 Hz. The resulting profiles for both cases, calculated using Darendeli (2001), are presented in Figure 6, illustrating the differences in $D_{min}$ across the four sites.

It is also essential to account for spatial variability in damping, as $D_{min}$ is not uniform across a site but varies with changes in soil type and stratigraphy. To incorporate spatial variability in damping, pseudo-3D $D_{min}$ models were constructed by extending and scaling the 1D $D_{min}$ profiles at each downhole array using the same procedure applied in scaling the 1D Vs profiles for developing the pseudo-3D Vs models (i.e., by stretching or compressing soil layer boundaries based on variations in $f_{0,H/V}$ obtained from H/V noise measurements at the surface). From these pseudo-3D $D_{min}$ models, $D_{min}$ cross-sections were extracted to match the orientations of the corresponding Vs cross-sections. Figure 7 presents the complete pseudo-3D $D_{min}$ model for DPDA, along with the $D_{min}$ cross-section extracted along the same 165° azimuth as the Vs cross-section (refer to Figure 3). Notably, the shallow glacial till layer in the southeastern portion of the pseudo-3D $D_{min}$ model (Figure 7a) and its expression in the 165° azimuth $D_{min}$ cross-section (Figure 7b) highlight the importance of accounting for spatial variability in $D_{min}$ when performing multi-dimensional GRAs.

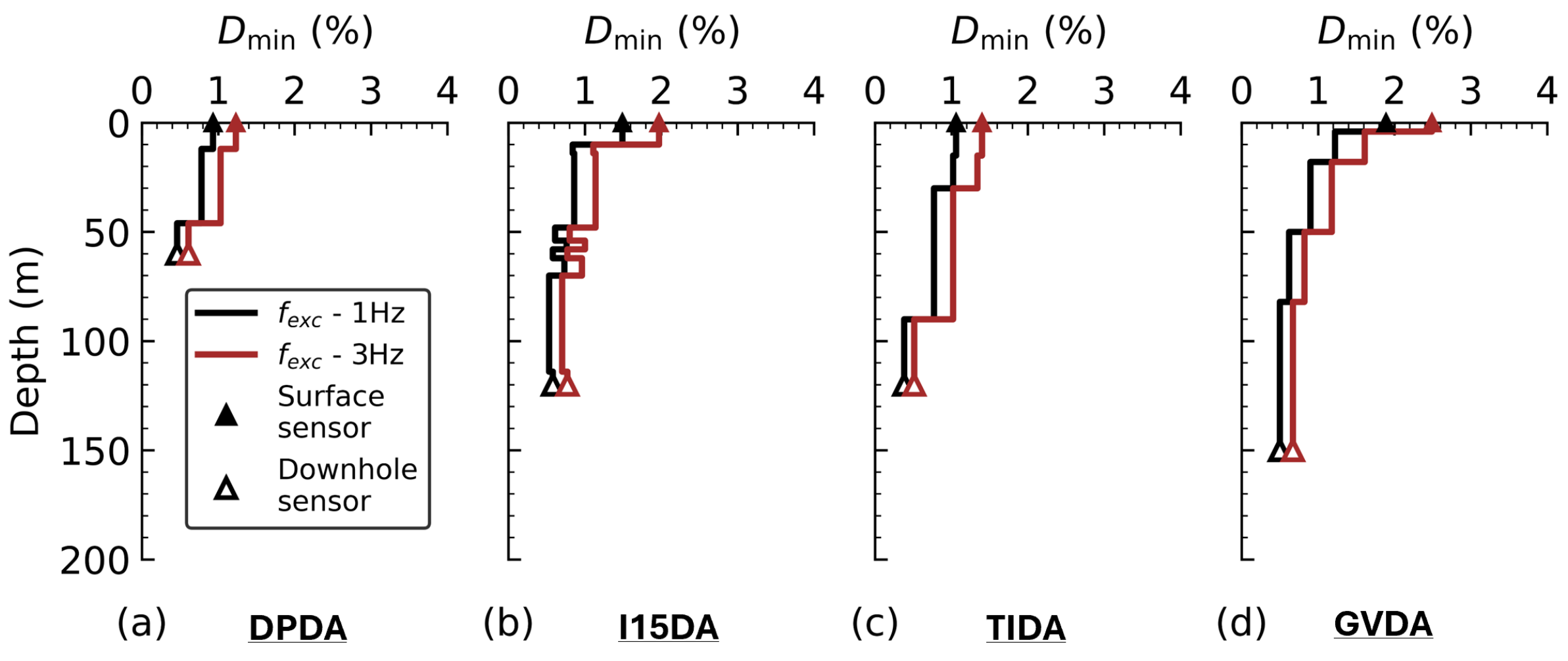

**Figure 6.** 1D $D_{min}$ profiles calculated using Darendeli (2001) for $f_{exc}$ of 1 Hz and 3 Hz, plotted against depth at: (a) DPDA, (b) I15DA, (c) TIDA, and (d) GVDA. Only the profiles corresponding to $f_{exc}$ = 3 Hz were adopted in this study.

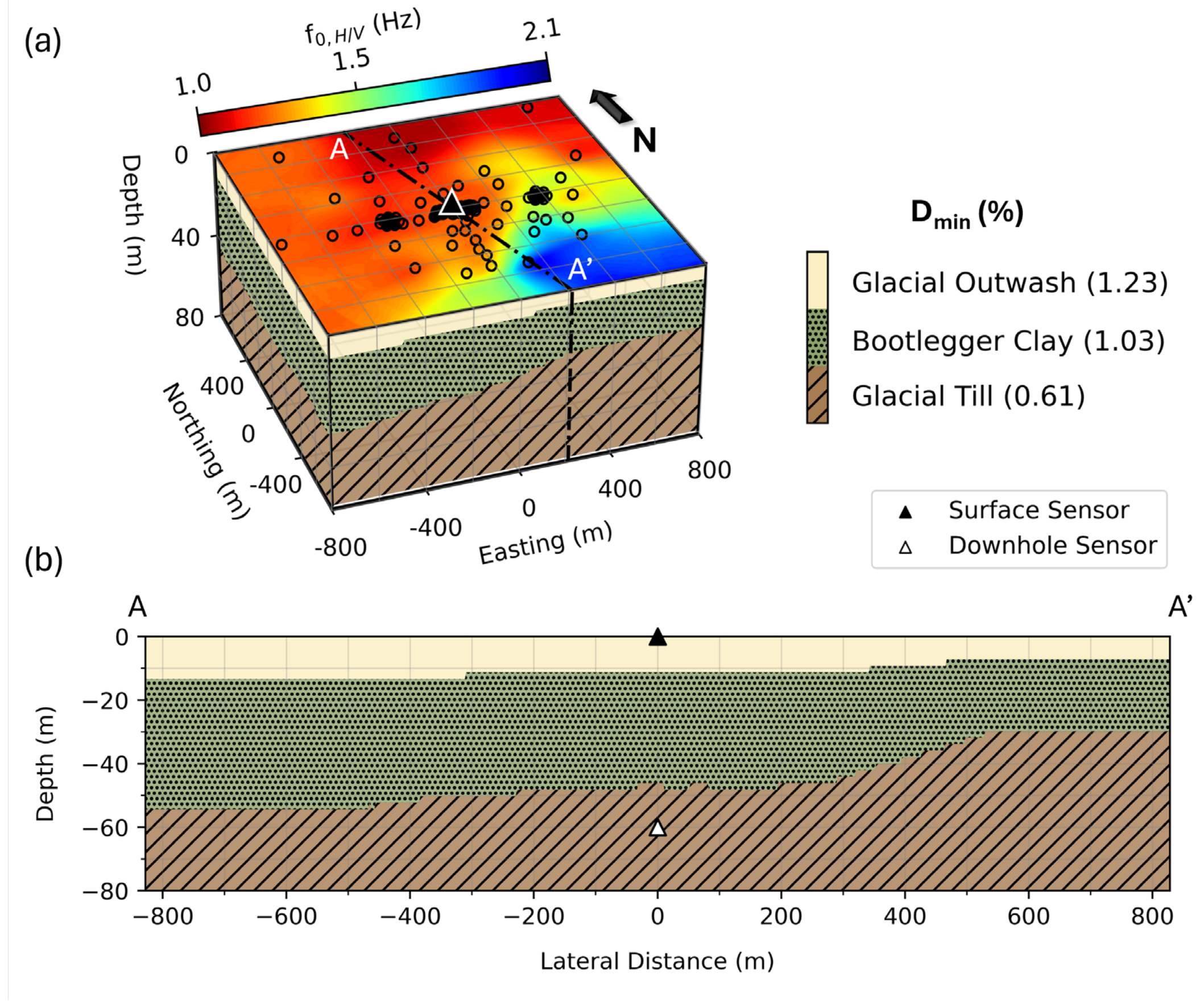

**Figure 7.** 3D variation in $D_{min}$ at DPDA. Shown are: (a) the pseudo-3D $D_{min}$ model, and (b) the $D_{min}$ cross-section extracted along the same azimuth (Az 165°) as the most heterogeneous Vs cross-section (AA') at DPDA. Solid and hollow triangular symbols denote the surface and downhole sensors, respectively.

### *5.3 Damping Multipliers*

As discussed earlier, in addition to intrinsic energy dissipation within soils, seismic waves also exhibit geometric and apparent damping, which cannot be captured using intrinsic damping values determined from laboratory testing. Geometric damping is relatively well understood, whereas apparent damping, which arises from wavefield phenomena such as wave scattering, reflection at interfaces, and mode conversions, is highly site-dependent and can complicate the application of damping ratios derived from laboratory measurements.

The pseudo-3D Vs model used in this study is site-specific and attempts to capture key variations in layer depths, including major impedance contrasts such as the soil–bedrock interface. During 2D and 3D GRA simulations, wave scattering, reflections, and mode conversions caused by such prominent impedance contrasts are explicitly captured (to a lesser degree in 2D than in 3D), thereby accounting for energy loss associated with apparent damping. However, because the pseudo-3D Vs model is primarily constrained by the Vs profile at the downhole array, it may not capture other sources of apparent damping, such as heterogeneity within each layer, subsurface lenses or irregular layer boundaries located away from the array that are not included in the simplified 3D model.

Furthermore, when input motion is incoherent, meaning that seismic waves do not arrive at the bedrock beneath the entire model in phase, additional scattering, reflections, and mode conversions can occur. Such incoherence can arise from variations in the wave path, irregularities in the subsurface, or differences in how seismic energy is radiated from the source. These effects increase the complexity of the incoming wavefield and increase apparent damping by redistributing energy across frequencies and directions before the waves even interact with the bedrock beneath the site. Due to the lack of information on such bedrock motions, the input motion in this study was applied uniformly at the base of the model. Although this is a simplifying assumption, it is justified by the inability to predict spatial incoherency prior to an actual earthquake.

In multi-dimensional GRAs, it is also common practice to apply excitation as a vertically propagating plane wave, allowing irregularities in bedrock and internal layer boundaries within the model to generate non-vertically propagating body and surface waves. However, when inclined waves arrive at the bedrock beneath the site, additional interference and reverberations can arise, further increasing wavefield complexity and contributing to apparent damping. While our research team has parametrically studied inclined input waves using 3D GRAs at the DPDA and I15DA sites and found that their effects on the TTFs are minimal for angles of incidence less than 15° (Dawadi and Cox, 2026; Dawadi et al., 2026), further studies are needed. In the absence of detailed information on such effects, the use of a vertically propagating uniform input remains a practical and reasonable assumption. Nonetheless, input motion incoherency, whether due to inclined incidence or other complexities arising outside the spatial extent of the model, can contribute to apparent damping and should be considered when interpreting site response results.

Therefore, to account for these unmodeled energy losses, 2D GRAs were also performed using inflated $D_{min}$ (i.e., $\varpi \times D_{min}$) values. These inflated $D_{min}$ values were calculated on a site-by-site basis to minimize the differences in the amplitudes of the $f_0$ peak of the analytically-calculated 1D TTF and the $LM_{ETF,EW\&NS}$ at each site, as done by Tao and Rathje (2019). However, It is important to clarify how this procedure should be interpreted. The $\varpi$ values were calibrated using 1D analytical TTFs, consistent with prior studies, because iterative calibration in multi-dimensional GRAs would have been computationally prohibitive.

Although it could be argued that the optimal $m$ value for 2D GRAs should be smaller since some apparent damping is already captured directly through simulation, our results showed only minor differences between the $f_0$ peak amplitudes of 1D and 2D TTFs prior to FAS smoothing. In addition, the energy dissipation simulated through explicitly accounting for spatial variability in 2D GRAs generally appeared as secondary peaks away from $f_0$, whereas the inflated $D_{min}$ primarily corrected the overestimation of the $f_0$ peak amplitude. For these reasons, we determined that using the $m$ values obtained from 1D calibration in our 2D GRAs was appropriate within our modeling approach.

At each site, a range of multipliers was tested, and the smallest value for which the amplitude of the TTF $f_0$ peak fell within the $LM_{ETF,EW\&NS} \pm 1\,\sigma_{ln,ETF,EW\&NS}$ bounds was selected as the optimal $m$ value. The procedure used for selecting the appropriate $m$ value at DPDA is illustrated in Figure 8, wherein a best-fitting value of 8 was determined. Similar determinations were made for all four downhole array sites. It can be noted from Figure 8 that while the amplitude of the 1D TTF obtained using 8 × $D_{min}$ matched that of the ETF at the $f_0$ peak, higher-mode peaks were significantly underpredicted, similar to findings from Hallal et al. (2022b). This suggests the need for investigating frequency-dependent damping formulations.

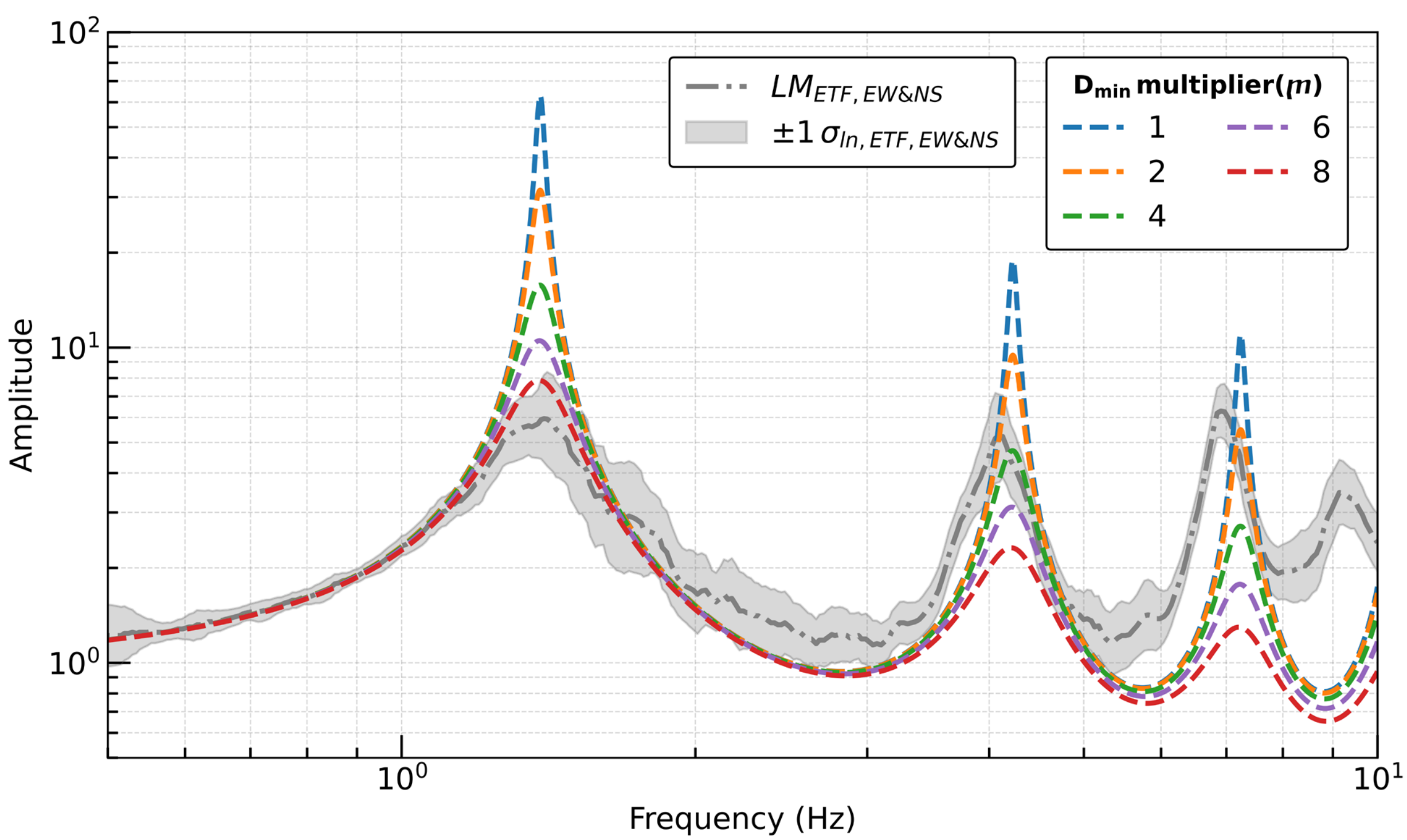

**Figure 8.** Selection of $m$ value at the DPDA site by scaling 1D TTFs obtained from an analytical closed-form solution to best match the $f_0$ peak amplitude of the combined ETF ($LM_{ETF,EW\&NS} \pm 1\,\sigma_{ln,ETF,EW\&NS}$).

For the four downhole array sites analyzed in this study, $m$ values ranged from 2 to 10. When plotted against the velocity contrast at each site, a strong negative correlation was observed between the magnitude of the velocity contrast and the required $m$ value, as shown in Figure 9. In this study, the velocity contrast is defined as the ratio of the Vs at the stiffest bedrock layer, indicated by the red circular symbol in Figures 9a–d, to the time-averaged Vs of all of the overlying soil layers above that depth, represented by the red dashed line in Figures 9a–d. At DPDA and I15DA, which have relatively low velocity contrasts of 4.83 and 2.93, respectively, high ɱ values of 8 and 10, respectively, were required. In contrast, GVDA and TIDA exhibit higher velocity contrasts of 7.84 and 9.47, respectively, and correspondingly required lower ɱ values of 3 and 2, respectively. This inverse relationship suggests that sites with lesser velocity contrasts experience lower empirical site amplification at $f_0$ than one might theoretically predict from a simple, 1D linear-viscoelastic analytical equation. This lower site amplification can only be modeled in 1D using an apparently higher $D_{min}$ that attempts to account for non-1D wave scattering effects. However, in multi-dimensional GRAs, one would hope that directly modeling spatial variability would diminish the need to use such significant empirical damping multiplier factors.

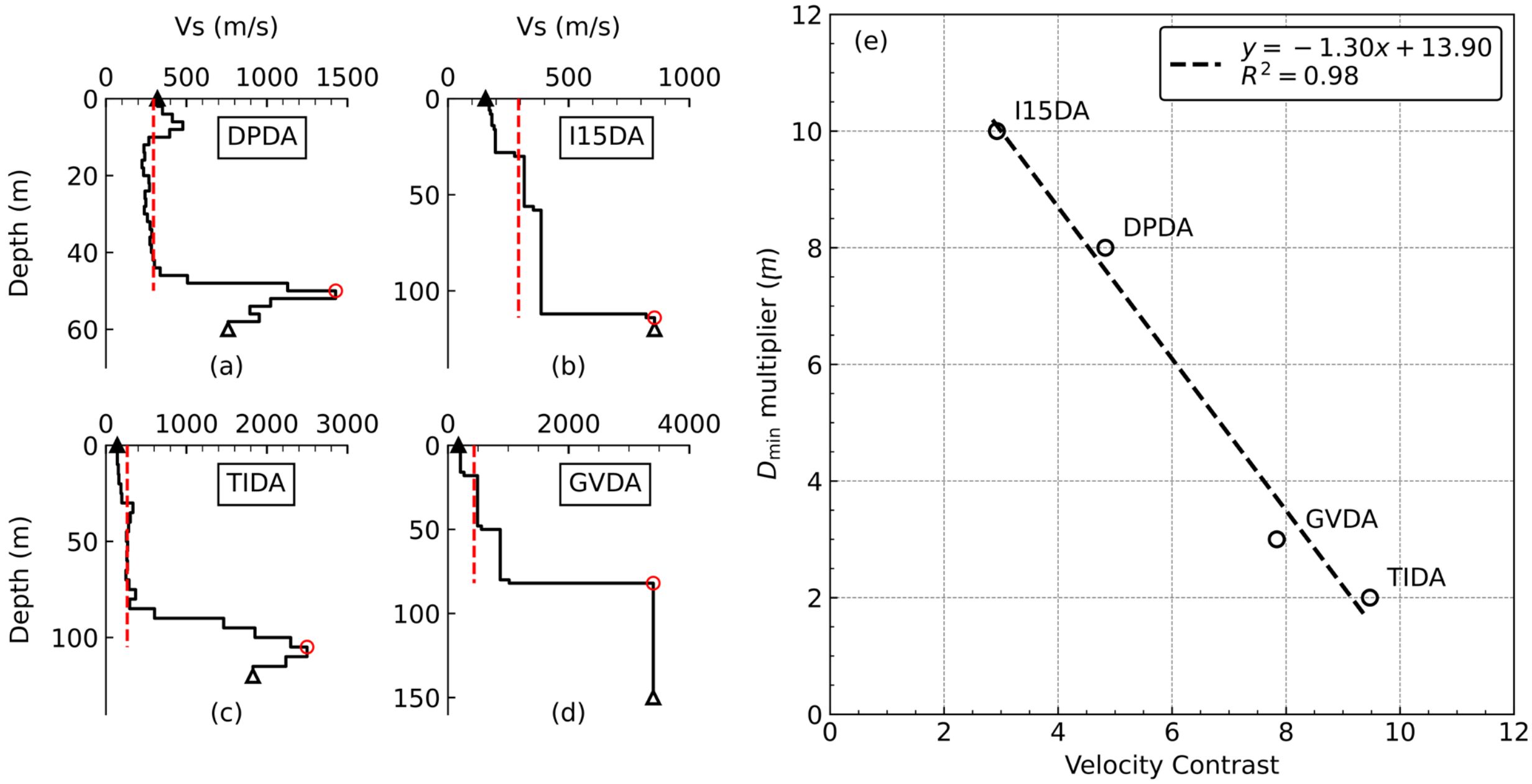

**Figure 9.** Relation between the velocity contrast at a site and the corresponding ɱ value required to match the $f_0$ peak amplitude in the 1D analytical TTF with that observed in the ETF at the site. Shown are the Vs profiles, highest Vs layer (red hollow circle symbol), and time-averaged Vs values for all layers above the top of the highest Vs layer (red dashed line) at: (a) DPDA, (b) I15DA, (c) TIDA, (d) GVDA; and (e) relationship between ɱ value and velocity contrast for each site. The ɱ values are based on Darendeli (2001) $D_{min}$ evaluated at $f_{exc}$ = 3 Hz.

### *5.4 Numerical Damping Formulations*

In this study, three distinct numerical damping formulations were implemented to evaluate their effectiveness in modeling energy dissipation during seismic wave propagation: Full Rayleigh Damping, Maxwell Damping, and Rayleigh Mass Damping. Each formulation offers unique advantages and limitations in terms of frequency dependence, computational efficiency, and alignment with target damping behavior. The following sections describe these formulations in detail and outline the rationale for their selection and application in GRAs for this study.

### *5.4.1 Full Rayleigh Damping*

Full Rayleigh Damping is a widely used method to model energy dissipation during vibration or seismic excitation (Hall, 2007). It assumes that the damping force is a linear combination of the mass and stiffness matrices of the system. The Rayleigh damping model is expressed using two coefficients, α and β, which represent the contributions of mass and stiffness to the total damping, respectively. The general equation for Rayleigh damping is:

$$\boldsymbol{C} = \alpha \boldsymbol{M} + \beta \boldsymbol{K} \tag{2}$$

where **C** is the damping matrix, **M** is the mass matrix, **K** is the stiffness matrix, α is the mass-proportional damping coefficient, and β is the stiffness-proportional damping coefficient. These coefficients are specified to achieve a target damping value over a specific frequency range of interest. The choice of α and β is critical for accurately modeling the wave dissipation behavior. A Full Rayleigh Damping schematic is shown in Figure 10. In this example, relatively constant damping of about 1% is obtained over a frequency range of approximately 2 – 8 Hz.

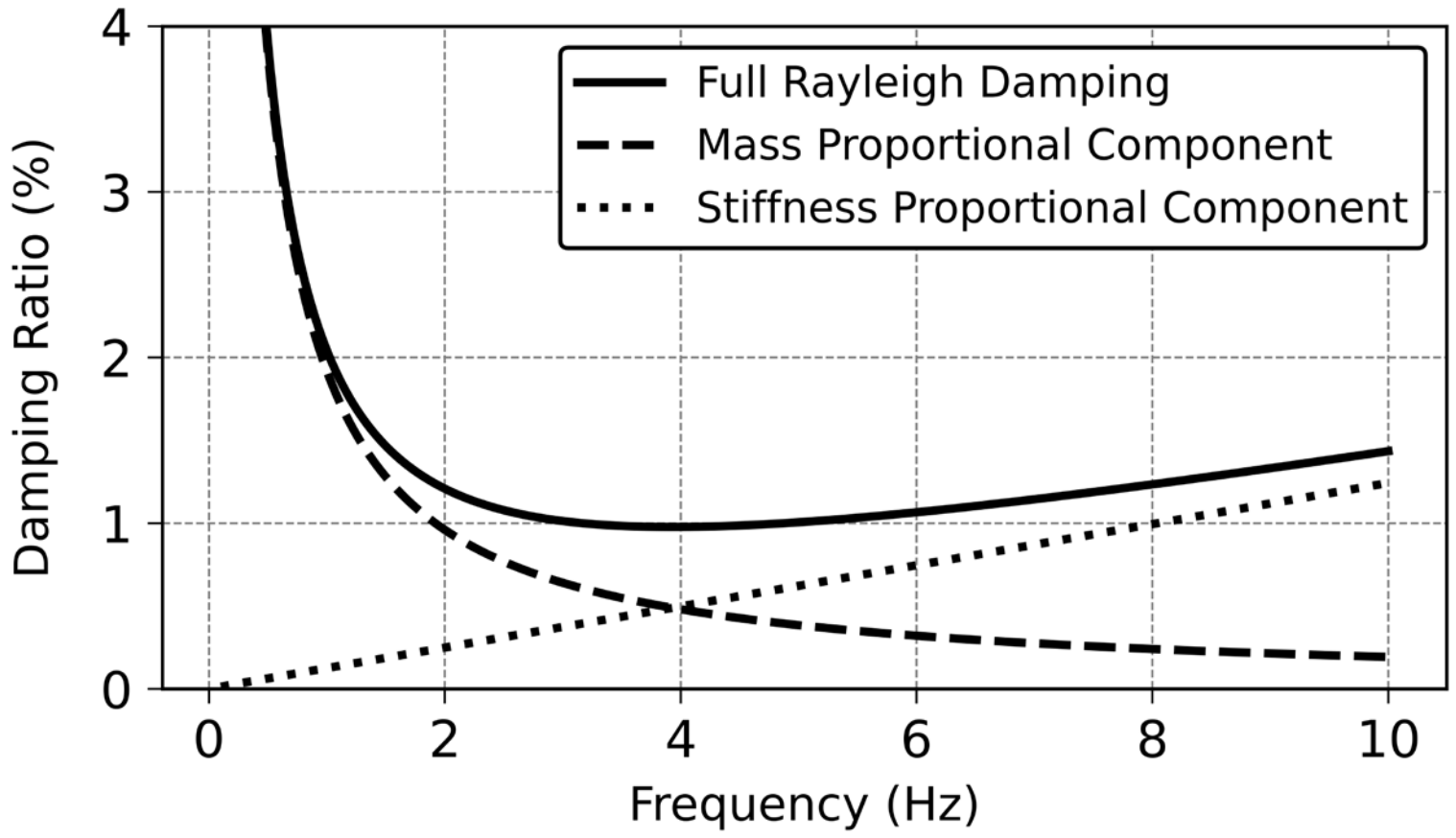

**Figure 10.** Full Rayleigh Damping scheme.

Full Rayleigh damping is commonly used to approximate frequency-independent damping over a specified frequency range, with the goal of producing a damping curve that remains as flat as possible. For the present study, we desire a flat damping curve between $f_0$ and $f_3$ of the ETFs at each site. Achieving this requires appropriate selection of mass and stiffness fitting parameters in numerical simulations. One common approach involves matching the target damping ratio at two selected frequencies, typically at the lower and upper bounds of the desired frequency range. This method, referred to here as the "low fit" (illustrated in Figure 11a), is widely used in practice. Another approach matches the minimum point of the Rayleigh damping curve to the target damping ratio. This is referred to here as the "high fit" (Figure 11b). In this study, an alternative approach based on root mean square (RMS) minimization was used to optimize the fit between the Rayleigh damping curve and a prescribed target damping ratio over the selected frequency range. In this approach, the damping ratio and frequency bounds are specified a priori, and the Rayleigh coefficients α and β are optimized to minimize the RMS difference between the resulting damping curve and the target. This optimized fitting method provides a more uniform representation of damping over the same frequency band, as shown in Figure 11c.

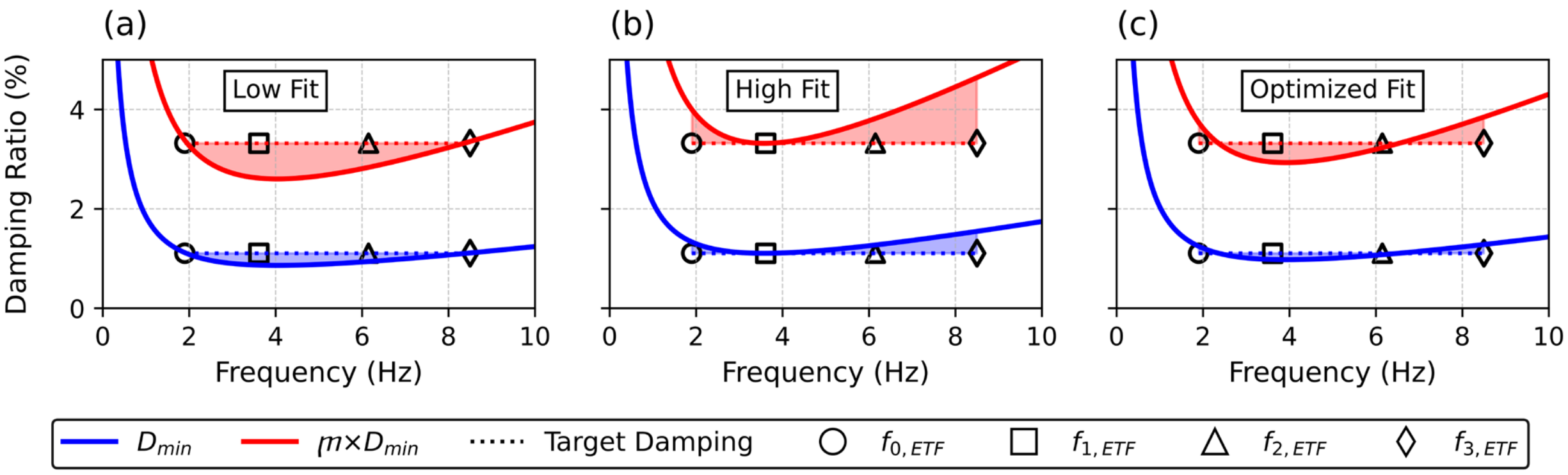

**Figure 11.** Full Rayleigh Damping formulation for two different target damping ratios (i.e., $D_{min}$ and ɱ × $D_{min}$). Shown are: (a) the low-fit method, (b) the high-fit method, and (c) an optimized-fit method using root mean square (RMS) minimization. In this illustration, the frequency peaks ($f_0$, $f_1$, etc.) correspond to the ETF modal frequencies at the GVDA site.

As illustrated in Figure 11, the differences between these fitting methods are minor at low damping ratios (e.g., $D_{min}$), but become increasingly significant as the target damping increases (e.g., ɱ × $D_{min}$). The Rayleigh damping formulation based on the low-fit method tends to underestimate damping within the frequency range between $f_0$ and $f_3$, while that based on the high-fit method tends to overestimate it. The optimized-fit approach offers a balanced solution, maintaining a relatively constant damping across the frequency range, even at higher target damping ratios. Notably, damping values at frequencies below the $f_0$ are significantly higher than the target, while damping values at frequencies above $f_3$ may also be elevated, depending on how flat the Full Rayleigh damping model remains outside the fitting bounds. This results in significant overdamping in numerical simulations for frequencies outside the fitting bounds.

#### *5.4.2 Maxwell Damping*

Bielak et al. (2011) and other researchers have shown that damping in time-domain analyses can be enhanced by replacing traditional stiffness-based damping with one or more Maxwell components, which consist of a spring in series with a dashpot. Each Maxwell component provides frequency-dependent damping, peaking at a specific frequency (Figure 12a). By combining multiple Maxwell components in parallel, damping can be made more consistent across a wider frequency range (i.e., frequency-independent). The use of additional components broadens the frequency range over which constant damping is maintained. For instance, Liu et al. (1976) utilized 12 Maxwell components to achieve nearly constant damping across four orders of magnitude in frequency. However, in more recent geophysical and seismological applications, typically only two or three Maxwell components are used (Fan et al., 2016; Bielak et al., 2011; Carcione et al., 1988). Unlike Full Rayleigh damping, Maxwell damping remains flat over the frequency range of interest, irrespective of the target damping ratio (Figure 12b). Maxwell damping is integrated into FLAC3D as one of the built-in damping formulations. For further details on its formulation and implementation, readers are referred to Dawson and Cheng (2021). However, this formulation has certain limitations, which will be discussed in greater detail later.

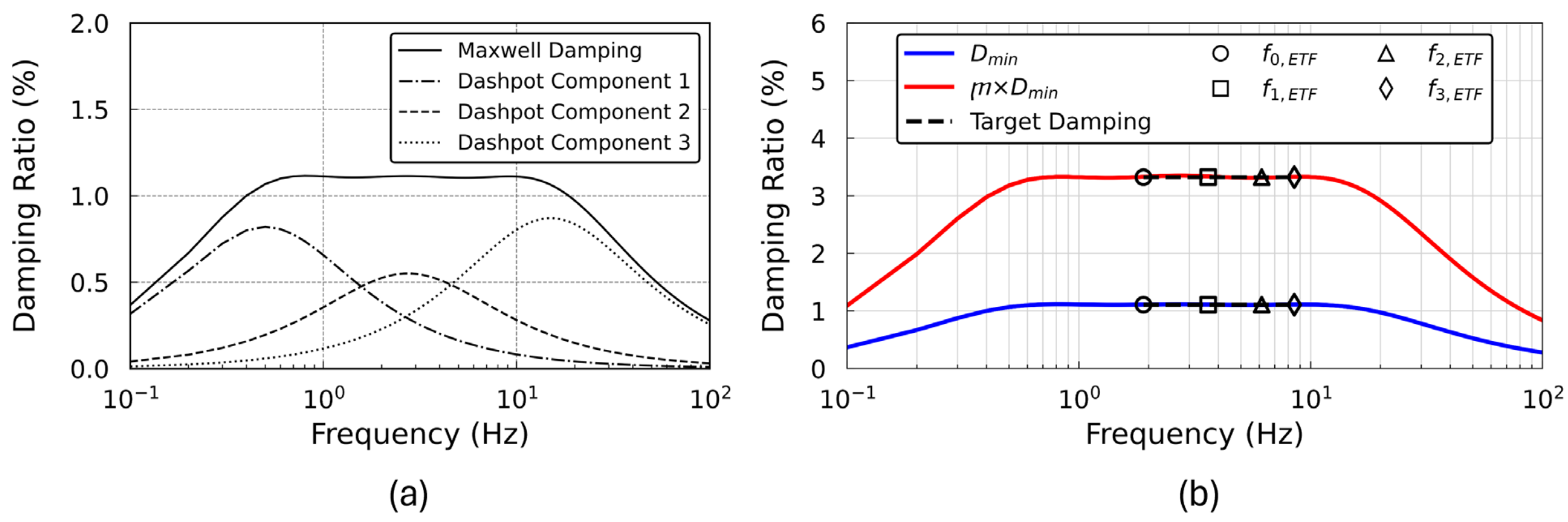

**Figure 12.** Maxwell Damping formulation. Shown are: (a) a schematic of a Maxwell damping curve obtained from three Maxwell dashpot components, and (b) an illustration showing the Maxwell damping scheme for two different target damping ratios ($D_{min}$ and $ɱ \times D_{min}$) at the GVDA site.

### *5.4.3 Rayleigh Mass Damping*

While Full Rayleigh and Maxwell damping schemes are designed to approximate frequency-independent damping over a specified frequency range, Rayleigh Mass damping is inherently frequency dependent, with higher damping at low frequencies and lower damping at high frequencies (refer to Figure 10). This section explores the rationale for using Rayleigh Mass damping in the present study. The mass-proportional component of Full Rayleigh Damping represents linear viscous dampers that connect the system's degrees of freedom to external support, while the stiffness-proportional component connects the degrees of freedom internally. In FLAC3D, the mass-proportional component can be applied independently, and this formulation is referred to as Rayleigh Mass Damping in this study. Hall (2007) highlights the potential limitations of Full Rayleigh damping, particularly the unrealistic damping of rigid body modes introduced by the mass-proportional term. Historically, the use of Full Rayleigh damping has been applied mainly in structural engineering, especially for building analysis, and its use as a standalone mass-only formulation remains limited. Despite its theoretical constraints, the motivation for using Rayleigh Mass damping in this study stems directly from empirical observations that frequency-independent damping models, including Full Rayleigh and Maxwell damping, consistently overdamp high frequencies, leading to systematic underprediction of high-frequency TTF peaks when compared with ETFs at the downhole array sites. This observation is discussed in greater detail in the 'Assessment of GRAs relative to ETFs' section, below.

Although using Rayleigh Mass damping alone may lack theoretical justification, it offers a pragmatic alternative that will be shown to work quite well for matching recorded ground motions at the borehole array sites considered herein. The physical mechanisms of wave attenuation in geomaterials are still not fully understood or easily modeled, and the ideal functional form of a numerical damping model that allows for better predictions relative to recorded ground motions remains undefined. As such, all damping models, including Full Rayleigh, Maxwell, and others, should be viewed as approximations to the "true" behavior with tunable parameters that may or may not carry physical meaning. This situation is analogous to nonlinear soil constitutive models, whose parameters are often calibrated outside strict physical ranges. Given this context, Rayleigh Mass damping was explored in the present study as a viable option. The availability of this formulation within FLAC3D allowed for its practical evaluation. A key advantage of this approach is its computational efficiency, particularly at higher damping ratios, where it outperforms Full Rayleigh damping in terms of analysis runtime. Details on computational performance are provided in a subsequent section.

Figure 13 illustrates the method used to apply Rayleigh Mass damping for a specified damping ratio. First, a Full Rayleigh damping curve was fitted to the target damping value using the RMS optimized-fit approach (Figure 13a). Then, the entire Full Rayleigh damping curve was shifted upward so that the mass-proportional component aligned with the target damping ratio at the ETF $f_0$ of the site (Figure 13b). Once this alignment was achieved, only the mass-proportional component was implemented as the damping mechanism in the numerical simulation.

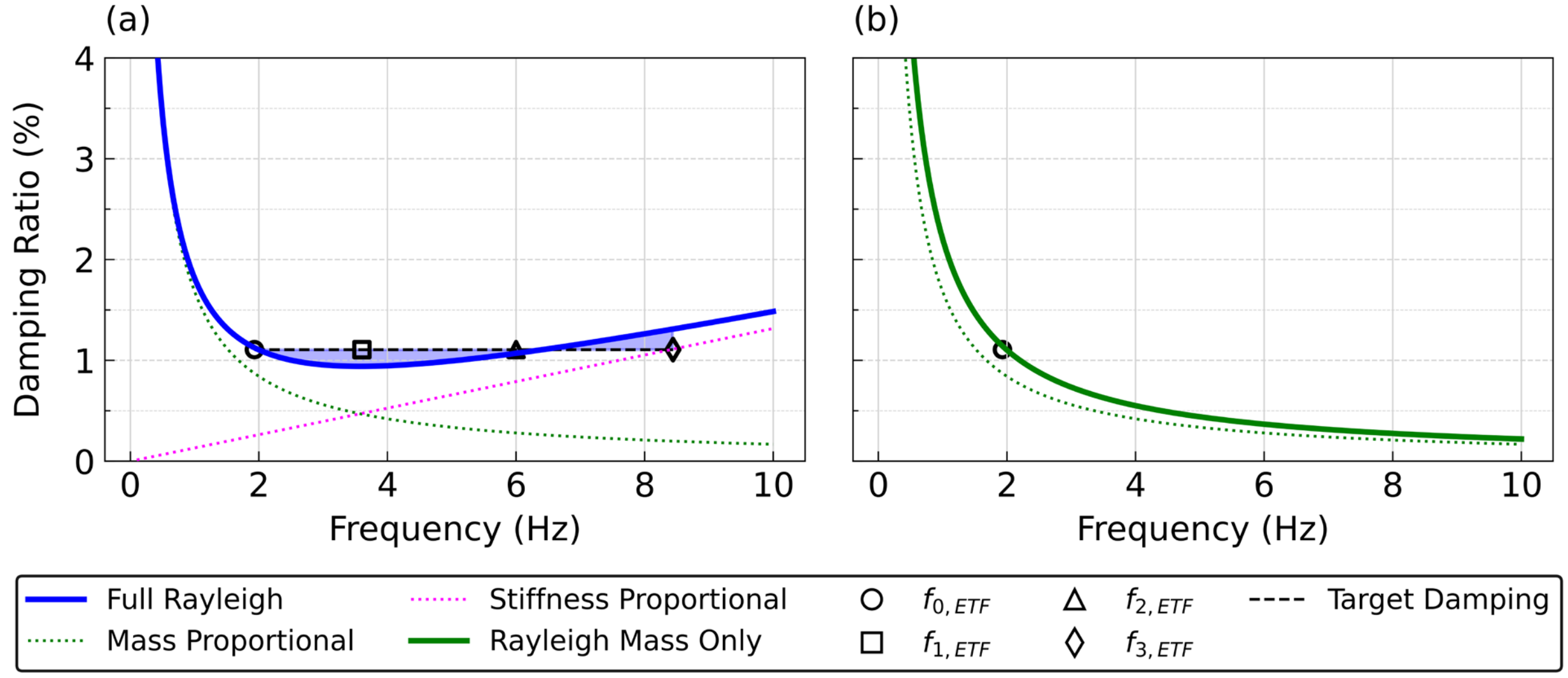

**Figure 13.** Formulation of Rayleigh Mass Damping. Shown are: (a) fitting a Full Rayleigh curve to the target damping ratio using the RMS optimized-fit approach, (b) shifting the Rayleigh Mass curve upward to match the target damping ratio at the ETF $f_0$ for the GVDA site.

### *5.5 Boundary Conditions*

Subsurface spatial variability leads to wave scattering, generating diffracted body and surface waves. Under such conditions, fixed boundaries are unsuitable because they produce strong reflections, requiring impractically large models to mitigate. Similarly, periodic or tied boundaries, which enforce identical motions on opposite sides, are incompatible with irregular models where material properties vary on either side of the model. To address these challenges, boundary conditions that minimize artificial wave reflections are essential. Therefore, free-field boundaries were applied along the vertical model sides to represent the dynamic response of the adjacent 1D soil columns and absorb outgoing scattered waves, i.e., the portion of the total wavefield not captured by the 1D response. In FLAC3D, these are implemented using free-field grids coupled to the main grid by Lysmer dashpots (Lysmer and Kuhlemeyer, 1969), which deform independently and transmit equivalent free-field forces to the model boundaries. This approach effectively simulates the infinite lateral extent of the model and prevents spurious side reflections. At the model base, a quiet boundary was used to absorb downgoing waves. Similar to free-field boundaries, quiet boundaries in FLAC3D are implemented using Lysmer dashpots that absorb outgoing energy, ensuring that wave energy leaving the model does not reflect back into the domain. Together, these boundary conditions allow the model to minimize artificial reflections, ensuring a more accurate simulation of wave propagation through laterally variable media.

### *5.6 Input Motion*

This study focuses on linear-viscoelastic GRAs, and we evaluate site response by computing TTFs as the ratio of the FAS of the simulated/recorded motion at the surface to that of the deepest sensor. The TTFs are later compared with ETFs derived from small-strain downhole recordings. In this framework (i.e., when comparing transfer functions), it is not necessary to use actual ground motions recorded at the base of the borehole as input motions. Instead, any input motion with sufficient energy across the frequency band of interest is suitable for analysis. Following common practice in prior studies (e.g., Hallal and Cox, 2023), we used Ricker wavelets as input motions, as they provide a well-controlled, broadband source ideal for evaluating frequency-dependent site response.

Selecting an input wavelet with adequate frequency content across the target range is essential in GRA simulations to accurately capture the site's dynamic behavior across various frequencies. In this study, we performed linear-viscoelastic numerical analyses using a composite input formed by combining Ricker wavelets with different central frequencies. This approach ensures sufficient energy coverage across the frequency range of interest, specifically, from the frequency corresponding to the half-amplitude point preceding the $f_0$ peak in $LM_{ETF,EW\&NS}$ to the half-amplitude point following the $f_3$ peak. The normalized energy spectral density (ESD) of individual Ricker wavelets with different center frequencies is shown in Figure 14 along with the $LM_{ETF,EW\&NS}$ at each site. ESD represents the distribution of energy per unit bandwidth for a given signal. As illustrated in Figure 14, a Ricker wavelet with a 1 Hz center frequency contains little energy above 2 Hz, whereas a wavelet centered at 5 Hz lacks sufficient energy below 2 Hz. This makes individual Ricker wavelets ineffective as standalone inputs for broadband excitation. To address this limitation, a broadband input was constructed by superimposing five Ricker wavelets with center frequencies logarithmically spaced between 0.5 Hz and 10 Hz (specifically, 0.50 Hz, 1.06 Hz, 2.24 Hz, 4.73 Hz, and 10.00 Hz). The resulting composite wavelet provides an even energy distribution across the 0.5–10 Hz range, with no frequency band receiving less than 50% of the peak spectral energy. This range encompasses both the $f_0$ peak and the first three higher-mode peaks identified in the ETFs at all downhole array sites considered in this study.

Since a quiet boundary was applied at the model base to suppress artificial reflections, the input motion was applied as stress. As quiet boundaries are implemented using dashpots, the consistent way to transmit seismic energy while simultaneously absorbing outgoing waves is through equivalent stress (force) loading.

## 6. Assessment of GRAs relative to ETFs

To evaluate how the different subsurface damping formulations and 3D Vs models influence site response predictions at each downhole array site, the input Ricker wavelet motions discussed above were propagated through the numerical models using FLAC3D, and synthetic ground motions were recorded at both the surface and deepest downhole accelerometer locations. These simulated ground motions were used to calculate TTFs as the ratio of the FAS of the simulated surface motion to that of the simulated motion at the deepest downhole sensor. These TTFs provide insight into predicted frequency-dependent amplification, showing how seismic energy is modified as it travels from the deepest sensor to the surface. To ensure accurate comparisons with ETFs, all FAS used in this study were smoothed using the Konno and Ohmachi (1998) method with a bandwidth coefficient (b-value) of 75, which is a more moderate smoothing compared to the b-value of 40 which is typically used for H/V processing (e.g., Molnar et al., 2022; Cox et al., 2020). It should be noted that both the ETFs and TTFs calculated in this study are assumed to be consistent with "within" transfer functions, meaning they account for both upgoing and downgoing seismic waves at the deepest subsurface sensor. The presence of downgoing waves in the recorded data was confirmed through deconvolution of the downhole acceleration time history from the surface motion, reinforcing the treatment of both ETFs and TTFs as within transfer functions. To ensure consistency with the ETFs, the TTFs were calculated using the simulated motions at the depth of the deepest downhole accelerometer, rather than from the input motion, which contains only upgoing waves. Consequently, all numerical cross-sections in this study were modeled to extend below the deepest downhole sensor at each site, with the input motion applied at the base.

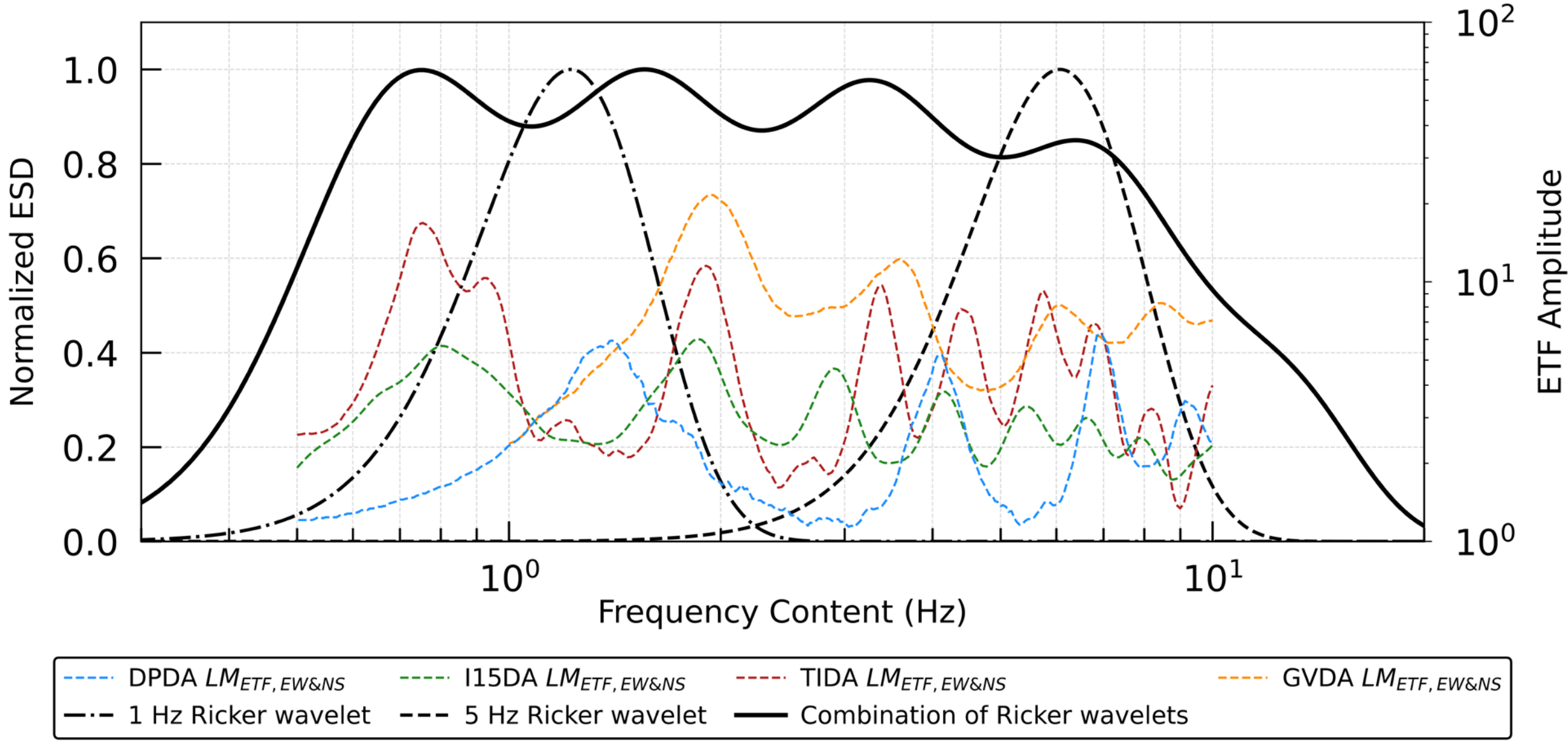

**Figure 14.** Normalized energy spectral density of input Ricker wavelets used for numerical simulation plotted against frequency range of interest. Also shown are the $LM_{ETF,EW\&NS}$ for the four downhole array sites as a means to visualize the need to excite input energy over a frequency range of approximately 0.5–10 Hz.

In the present study, eight different GRAs were performed at each downhole array site, as summarized in Table 4. At each site, site response was evaluated using three different damping formulations: Full Rayleigh damping, Maxwell damping, and Rayleigh Mass damping. For each damping formulation, two levels of damping were applied: the conventional small-strain damping ratio, $D_{min}$, calculated using Darendeli (2001), and the inflated $D_{min}$ value, $m \times D_{min}$. These six cases were then compared with one another and with 1D analytical linear-viscoelastic TTFs computed using both $D_{min}$ and $m \times D_{min}$. It should be noted that the Kelvin–Voigt damping model (Voigt, 1892; Thomson, 1865) was adopted in calculating the 1D TTFs using the analytical solution. The Kelvin–Voigt model represents viscoelastic behavior using a spring and dashpot in parallel, where stress is proportional to both strain and strain rate. This formulation results in frequency-dependent damping behavior, with the damping ratio increasing with frequency. However, it is commonly modified to achieve frequency independent damping by implementing an equivalent viscosity that is inversely proportional to frequency (Kramer & Stewart, 2025). Finally, all eight cases were evaluated against the ETFs at the respective sites.

**Table 4.** Summary of GRA cases conducted at each downhole array during this study

| Cases | GRA | Damping Ratio | Damping Formulation |
|---|---|---|---|
| Case 1 | 1D Analytical | $D_{min}$ | Kelvin-Voigt |
| Case 2 | 1D Analytical | $\mathscr{m} \times D_{min}$ | Kelvin-Voigt |
| Case 3 | 2D Numerical | $D_{min}$ | Full Rayleigh |
| Case 4 | 2D Numerical | $\mathscr{m} \times D_{min}$ | Full Rayleigh |
| Case 5 | 2D Numerical | $D_{min}$ | Maxwell |
| Case 6 | 2D Numerical | $\mathscr{m} \times D_{min}$ | Maxwell |
| Case 7 | 2D Numerical | $D_{min}$ | Rayleigh Mass |
| Case 8 | 2D Numerical | $\mathscr{m} \times D_{min}$ | Rayleigh Mass |

To ensure consistency in multi-dimensional effects between simulations and observations, 2D TTFs were compared against the rotated ETFs ($LM_{ETF,Rotated} \pm 1\,\sigma_{ln,ETF,Rotated}$), as described in Section 4. Although comparison of 1D TTFs with rotated ETFs is not strictly appropriate, the 1D TTFs are included for reference to provide context for the 2D GRA results. The evaluation of the eight TTFs relative to the rotated ETFs was performed using both qualitative and quantitative assessments. The qualitative assessment is based on by-eye judgement of how closely the peaks and troughs in the TTFs align with the ETFs. The quantitative assessment evaluates the predictive accuracy of the TTFs by comparing them to the ETFs using two evaluation metrics: the Pearson correlation coefficient ($r$) and the transfer function misfit ($m_{TF}$). The correlation coefficient quantifies how closely the TTFs align with the median ETF, with higher values indicating better predictions. A perfect match yields $r$ = 1.0, and values above 0.6 are considered good (Thompson et al. 2009). The $m_{TF}$ measures the average deviation of the TTF from the median ETF in terms of standard deviations. While there are no definitive guidelines for acceptable $m_{TF}$ values due to its dependence on $\sigma_{ln,ETF}$, which varies frequency-by-frequency and across sites, lower $m_{TF}$ values are preferred, with 1.0 indicating that the TTF is within one standard deviation of the median ETF across the considered frequencies. The equations for calculating the $r$ and $m_{TF}$ values can be found in Teague et al. (2018). In this study, $r$ and $m_{TF}$ values were calculated over the frequency range extending from the half-amplitude point preceding the $f_0$ peak in $LM_{ETF,Rotated}$ to the half-amplitude point following the $f_3$ peak, or up to 10 Hz, whichever was smaller, following the procedures outlined in Hallal and Cox (2021b). These frequency limits are indicated with red dash-dot lines in Figures 15-18 for each site.

### *6.1 TTFs at DPDA*

The GRA results at the DPDA site (Figure 15) provide a detailed assessment of the performance of different damping formulations under both conventional and inflated $D_{min}$ values. When conventional $D_{min}$ was applied, regardless of the numerical damping formulation, the TTFs consistently overpredicted the amplitude of the $f_0$ peak (Figure 15a). The $f_1$ peak was also somewhat overestimated, while the $f_2$ peak showed reasonable agreement with the ETFs. This pattern was consistent across all three 2D damping formulations (Full Rayleigh, Maxwell, and Rayleigh Mass) as well as the 1D analytical solution, indicating that the $f_0$ peak tends to be systematically overpredicted when using conventional $D_{min}$, regardless of the GRA modeling approach. Since the FAS used to calculate the 2D numerical TTFs were smoothed using the

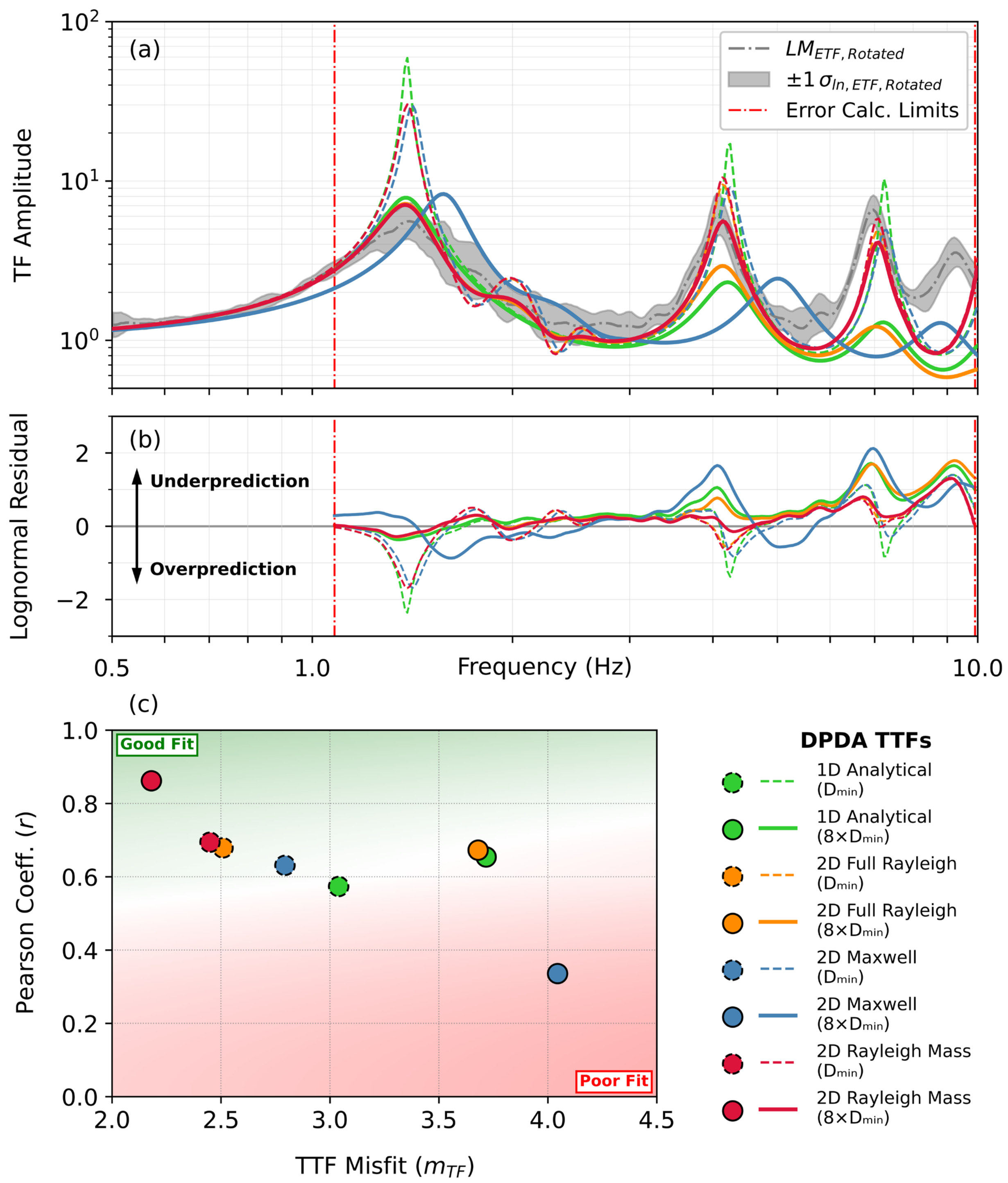

**Figure 15.** Site response predictions obtained from eight GRA cases using different damping formulations at DPDA. Shown are: (a) comparison between simulated TTFs and ETF, (b) lognormal residuals between the TTFs and ETF showing under/over prediction, and (c) Pearson correlation coefficient ($r$) and TTF misfit ($m_{TF}$) values associated with each case. Higher r and lower $m_{TF}$ indicate closer agreement with the ETF.

Konno and Ohmachi (1998) method, the amplitude differences between the 1D and 2D TTFs arise partly from this smoothing and partly from the ability of the 2D GRAs to capture spatial variability and subsequent wave scattering, as reflected by the secondary peaks observed near 2 Hz.

When the inflated $D_{min}$ ($8 \times D_{min}$) was applied, the results diverged more significantly. Full Rayleigh and Maxwell damping both captured the $f_0$ peak amplitude well, but severely underpredicted the higher-mode peaks due to overdamping. Maxwell damping also introduced a systematic shift of all three peaks toward higher frequencies, further degrading the fit. In contrast, Rayleigh Mass damping provided the closest overall match to the ETF. It not only reproduced the $f_0$ peak amplitude with strong agreement but also maintained close alignment with the $f_1$ peak, while only slightly underpredicting the $f_2$ peak. Compared to the 1D analytical solution at $8 \times D_{min}$, which essentially suppressed higher-mode peaks, the 2D Rayleigh Mass case preserved them more effectively.

The lognormal residuals between the median ETF and the TTFs (Figure 15b) highlight these differences more clearly. At $D_{min}$, residuals at $f_0$ were predominantly negative across all formulations, confirming systematic overprediction. At $8 \times D_{min}$, residuals for Full Rayleigh and Maxwell damping at $f_2$ and $f_3$ reached values as high as 2, reflecting strong underprediction. Rayleigh Mass damping limited residual magnitudes to within 1 across the full frequency range and to less than 0.3 up to 5 Hz, indicating much closer agreement with the ETF.

The performance of the damping formulations was also quantified using $r$ and $m_{TF}$ values, as shown in Figure 15c. A higher $r$ indicates stronger agreement in spectral shape between the TTF and ETF, while a lower $m_{TF}$ denotes a smaller amplitude misfit. Among the inflated $D_{min}$ cases, Rayleigh Mass damping achieved the highest $r$ (0.86) and the lowest $m_{TF}$ (2.18). This represents an improvement of 28% in correlation compared to Full Rayleigh ($r$ = 0.67) and 153% compared to Maxwell ($r$ = 0.34), while simultaneously reducing misfit by 41% relative to Full Rayleigh ($m_{TF}$ = 3.68) and 46% relative to Maxwell ($m_{TF}$ = 4.04). Conversely, Maxwell damping with inflated $D_{min}$ performed the worst, yielding both the lowest correlation and the largest misfit, consistent with the frequency shifts observed in Figure 15a.

### *6.2 TTFs at I15DA*

The results for all eight GRA cases at the I15DA site are presented in Figure 16. When conventional $D_{min}$ was applied, the TTFs predicted significantly elevated peaks at $f_0$ and at higher modes (Figure 16a), similar to the findings at DPDA. When inflated $D_{min}$ ($10 \times D_{min}$) was applied, the 1D analytical solution and 2D TTFs with Full Rayleigh and Maxwell damping all captured the $f_0$ peak reasonably well but substantially underpredicted higher modes due to overdamping, with Maxwell damping additionally introducing frequency shifts that became more pronounced at higher modes. In contrast, Rayleigh Mass damping yielded the closest overall match to the ETF under inflated $D_{min}$, accurately reproducing the $f_0$ peak amplitude while preserving the amplitudes of the higher-mode peaks without the severe amplitude reductions seen with other formulations.

The lognormal residuals (Figure 16b) also illustrate these differences. At $D_{min}$, all formulations exhibited negative residuals at $f_0$, consistent with systematic overprediction. At $10 \times D_{min}$, residuals for Full Rayleigh and Maxwell damping at the higher modes exceeded 1 and even approached 2 near 4 Hz, confirming strong underprediction. For Maxwell damping, peak shifts were also reflected in its residual distribution. On the other hand, Rayleigh Mass damping maintained residual magnitudes within 1.1 across the entire frequency range, with values below 0.5 up to about 2.3 Hz, indicating much closer agreement with the ETF.

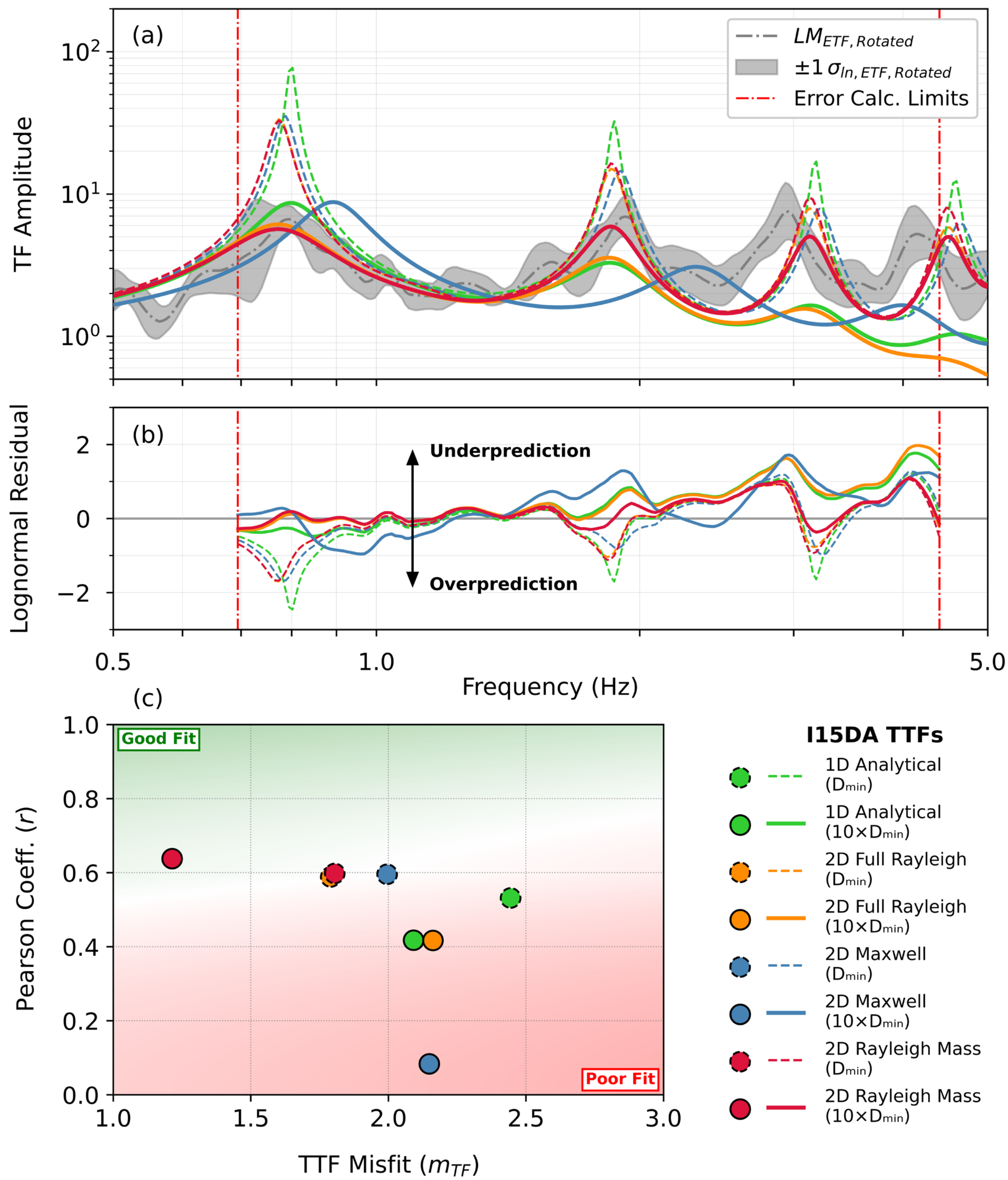

**Figure 16.** Site response predictions obtained from eight GRA cases using different damping formulations at I15DA. Shown are: (a) comparison between simulated TTFs and ETF, (b) lognormal residuals between the TTFs and ETF showing under/over prediction, and (c) Pearson correlation coefficient ($r$) and TTF misfit ($m_{TF}$) values associated with each case. Higher $r$ and lower $m_{TF}$ indicate closer agreement with the ETF.

The quantitative results are summarized in Figure 16c. Among the inflated $D_{min}$ cases, Rayleigh Mass damping achieved the best overall performance with the highest $r$ (0.64) and the lowest $m_{TF}$ (1.21). This corresponds to a 53% improvement in correlation relative to Full Rayleigh ($r$ = 0.42) and more than six times higher than Maxwell ($r$ = 0.08). In terms of misfit, Rayleigh Mass reduced $m_{TF}$ by 44% compared to Full Rayleigh ($m_{TF}$ =2.16) and 44% compared to Maxwell ($m_{TF}$ = 2.15). By contrast, Maxwell damping with inflated $D_{min}$ gave the poorest performance, combining the lowest correlation with high misfit, which is consistent with the frequency shifts observed in Figure 16a.

### *6.3 TTFs at TIDA*

The results from all eight GRA cases at the TIDA site are shown in Figure 17. The 1D analytical TTF using $D_{min}$ slightly overpredicted peak amplitudes at $f_0$ and $f_1$ (Figure 17a). However, compared to DPDA and I15DA, the 1D TTF overprediction of the ETF amplitudes at TIDA was less pronounced. With inflated $D_{min}$ (2 × $D_{min}$), the $f_0$ peak amplitudes aligned more closely with the ETF, but the 1D analysis was unable to capture the broadened $f_0$ peak that included a secondary peak just after the $f_0$ in the ETF (between 0.9 and 1.0 Hz). Hallal and Cox (2023) attributed this secondary peak to waves scattering from dipping bedrock that outcrops at Yerba Buena Island, located about 1 km from the borehole array. Their 2D GRA simulations reproduced the secondary peak, although the frequency alignment was not exact, and their analyses were limited to resolving frequencies up to $f_1$. In contrast, the present study used smaller element sizes, which extended frequency resolution and allowed ETF peaks to be identified up to $f_3$.

As shown in Figure 17, the 2D GRAs reproduced the spectral shape around $f_0$ more effectively than the 1D cases for both $D_{min}$ and 2 × $D_{min}$. Most damping formulations matched the observed response across the full frequency range, except for Maxwell damping with inflated $D_{min}$, which introduced frequency shifts at higher modes that degraded performance. At $f_0$, both Full Rayleigh and Rayleigh Mass damping with $D_{min}$ and 2 × $D_{min}$ performed comparably well. However, differences became more apparent at higher modes. At $f_1$, the Full Rayleigh peak was slightly lower than the Rayleigh Mass peak, though both remained very close to the ±1 $\sigma_{ln,ETF}$ bounds. At $f_2$, Rayleigh Mass with both $D_{min}$ and 2 × $D_{min}$, and Full Rayleigh with $D_{min}$, predicted peaks within the bounds, whereas Full Rayleigh with 2 × $D_{min}$ underpredicted the amplitude. At $f_3$, Rayleigh Mass with both $D_{min}$ and 2 × $D_{min}$ continued to produce peaks very close to the bounds, while Full Rayleigh underpredicted in both cases, with the 2 × $D_{min}$ case showing significant underestimation.

The lognormal residuals (Figure 17b) further illustrate the superiority of Rayleigh Mass damping, particularly at $f_2$ and $f_3$. In all cases other than Maxwell damping with 2 × $D_{min}$, residuals remained within ±1.0 across the entire frequency range, deviating only near the secondary peak close to $f_0$ in the ETF (around 0.9 Hz). While the Maxwell case with inflated $D_{min}$ was the main outlier, exhibiting residuals close to 1.0 at all three higher-mode peaks, Rayleigh Mass damping maintained residuals below 0.5 for frequencies above 1 Hz.

The statistical comparison in Figure 17c confirms the stronger performance of Rayleigh Mass damping. With inflated $D_{min}$, it achieved the highest $r$ (0.83) and the lowest $m_{TF}$ (1.48). This represents a 4% improvement in correlation compared to Full Rayleigh ($r$ = 0.79) and a 24% improvement compared to Maxwell ($r$ = 0.67). In terms of misfit, Rayleigh Mass reduced $m_{TF}$ by 27% relative to Full Rayleigh ($m_{TF}$ = 2.03) and 49% relative to Maxwell ($m_{TF}$ = 2.89). The poorest performance was once again observed for Maxwell damping with inflated $D_{min}$.

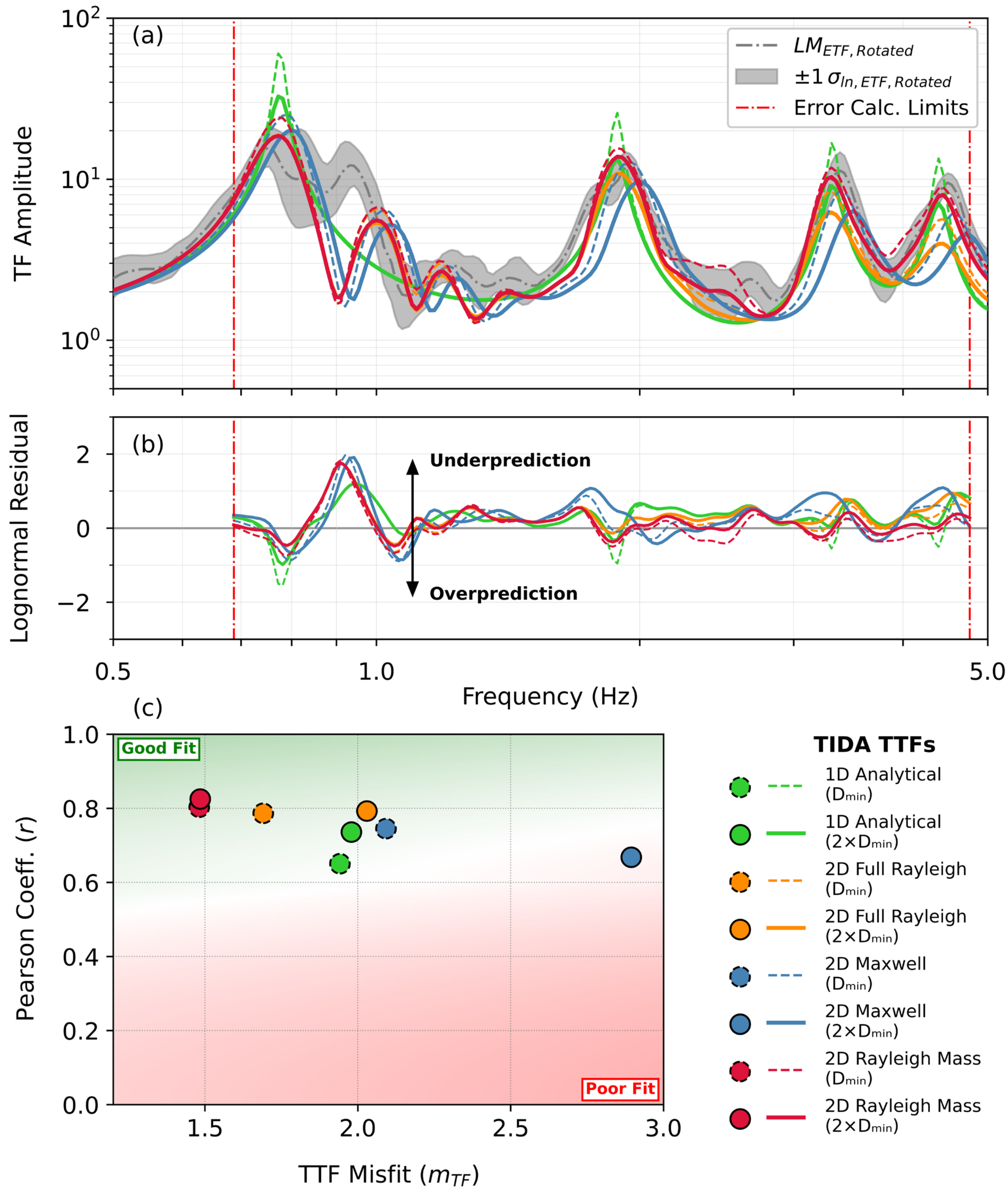

**Figure 17.** Site response predictions obtained from eight GRA cases using different damping formulations at TIDA. Shown are: (a) comparison between simulated TTFs and ETF, (b) lognormal residuals between the TTFs and ETF showing under/over prediction, and (c) Pearson correlation coefficient ($r$) and TTF misfit ($m_{TF}$) values associated with each case. Higher $r$ and lower $m_{TF}$ indicate closer agreement with the ETF.

Compared to DPDA and I15DA, the site response predictions at TIDA were already in good agreement without using any ɱ value, with all cases achieving $r$ values greater than 0.6 and TTF misfits below 2.1, except for the Maxwell damping with inflated damping, which was affected by frequency shifts. While the use of ɱ = 2 in the 2D GRAs at TIDA provided some improvement in peak amplitude agreement, the overall impact was limited. This is likely the result of two factors. First, TIDA has the strongest velocity contrast of all downhole array sites considered in this study (refer to Figure 9). Sites with stronger velocity contrasts tend to have higher ETF amplitudes that agree better with TTF amplitudes. Second, seismic energy at TIDA is primarily redistributed through interference caused by scattering from the dipping bedrock near Yerba Buena Island, which is well represented in the TIDA pseudo-3D Vs model (refer to Figure 2c) and is manifested by the prominent secondary peaks near $f_0$ in the TTFs. Since apparent damping effects resulting from energy redistribution away from $f_0$ are already captured by the modeled spatial variability in 2D, making additional damping adjustments was less critical for achieving more accurate results.

### *6.4 TTFs at GVDA*

The GRA results for the GVDA site are shown in Figure 18. The 1D analytical TTFs displayed noticeable overprediction of amplitudes at all four modes (Figure 18a), though the bias was less pronounced compared to DPDA and I15DA. The largest discrepancies between 1D and 2D responses occurred around $f_1$: in the 1D case, the peak appeared narrow and sharply amplified, whereas in the 2D analyses it broadened due to wave reflections and interference, producing secondary peaks. Similar secondary peaks also appeared at $f_2$ and $f_3$, reflecting more complex wavefield interactions in the 2D response, which match the general flattening of the ETF at these higher modes. The higher frequency wave scattering can be attributed to the abruptly dipping, hard, granitic bedrock with Vs > 3400 m/s that lies at a depth of about 80 m at the downhole array and outcrops towards the northwest (refer to Figure 2d). This dipping bedrock generates strong wave reflections, which broaden the response and produce secondary peaks in the transfer functions, while flattening the high-frequency response.

At GVDA, a ɱ value of 3 was determined through calibration. When 3 × $D_{min}$ was applied in the 1D analytical case, it significantly improved amplitude agreement at the $f_0$ and $f_1$ peaks of the ETF, as seen in Figure 18a. In contrast, the effect of ɱ value in the 2D GRAs was minimal. Seismic waves propagating upward at GVDA are strongly deflected and scattered by the dipping granite layer, producing interference and reverberations that generate secondary peaks around $f_1$, $f_2$, and $f_3$ in the 2D TTFs. Since this spatial variability is already represented in the pseudo-3D Vs model and captured in the 2D GRAs, additional empirical damping contributed little beyond the attenuation captured by modeled spatial variability. While the secondary peaks helped improve agreement with the ETF at $f_0$ and $f_1$, they aligned less closely at $f_2$ and $f_3$.

The lognormal residuals (Figure 18b) further illustrate these observations. When using $D_{min}$, residuals were moderately negative at $f_0$, confirming overprediction. Inflated $D_{min}$ in the 1D case reduced these residuals, while the 2D formulations produced mixed results. Rayleigh Mass damping maintained residuals comparable to Full Rayleigh throughout much of the frequency range, though its performance at $f_2$ and $f_3$ was not as good because it failed to damp the highest frequencies enough to capture the extreme flattening of the ETF.

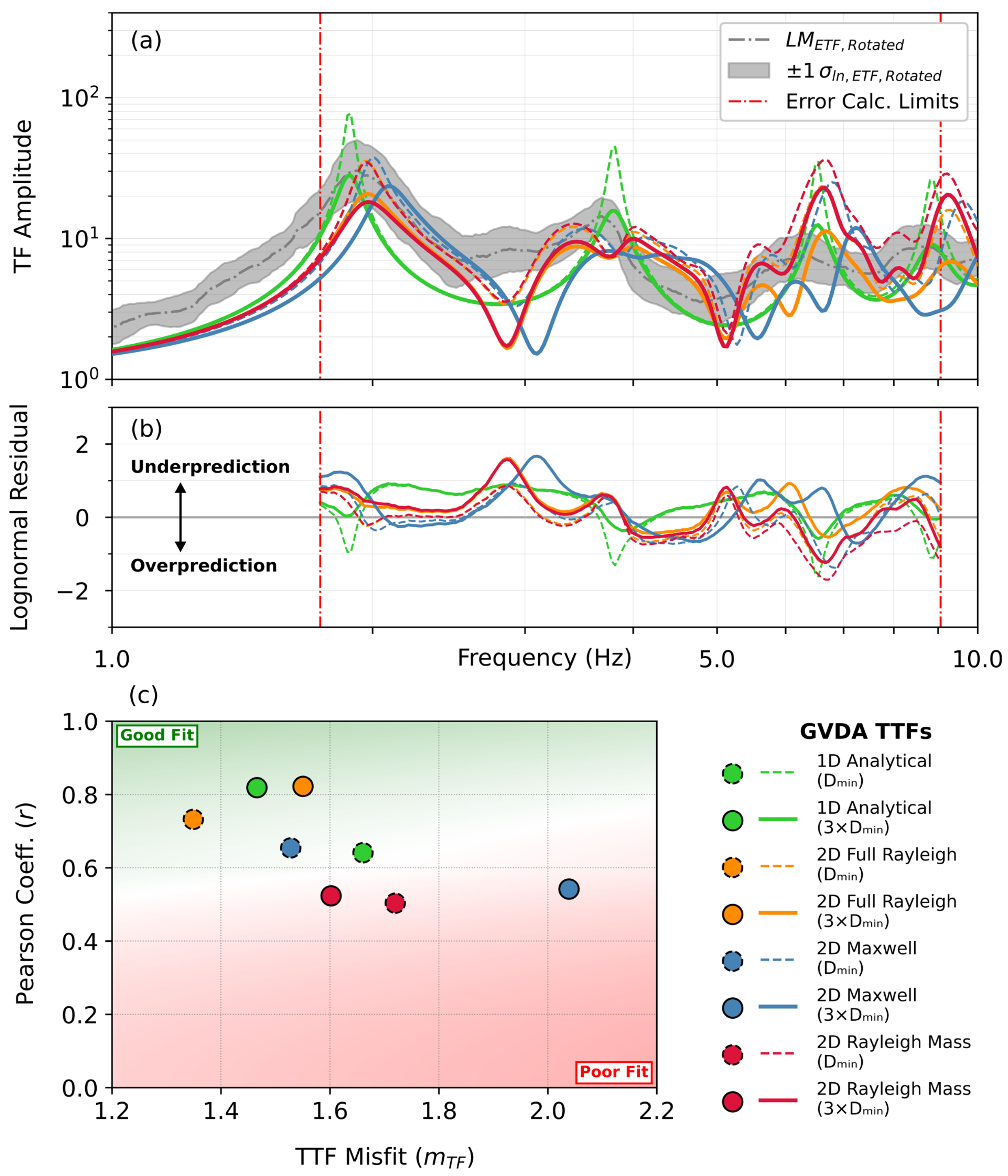

**Figure 18.** Site response predictions obtained from eight GRA cases using different damping formulations at GVDA. Shown are: (a) comparison between simulated TTFs and ETF, (b) lognormal residuals between the TTFs and ETF showing under/over prediction, and (c) Pearson correlation coefficient ($r$) and TTF misfit ($m_{TF}$) values associated with each case. Higher $r$ and lower $m_{TF}$ indicate closer agreement with the ETF.

The quantitative assessment in Figure 18c shows that Full Rayleigh damping with $D_{min}$ produced the lowest $m_{TF}$ (1.35), while Full Rayleigh with inflated $D_{min}$ achieved the highest $r$ (0.82). This is the only case presented herein where the lowest $m_{TF}$ and highest $r$ values didn't consistently come from an inflated $D_{min}$ value and Rayleigh Mass damping formulation. In fact, according to $r$ values, the Rayleigh Mass damping formulation performed the worst (i.e., lowest $r$ values). These low $r$ values are particularly influenced by overpredicted amplitudes at the higher-mode peaks, which are significantly flatter at GVDA than at the other borehole array sites. However, according to $m_{TF}$ values, the Maxwell damping formulation with inflated $D_{min}$ performed the worst due to shifts in the modal frequencies. To further complicate quantitative evaluation, Rayleigh Mass damping with inflated $D_{min}$ ($m_{TF}$ = 1.60) performed comparably to Full Rayleigh damping with inflated $D_{min}$ ($m_{TF}$ = 1.55) in terms of misfit. As noted by Teague et al. (2018), quantitative statistical metrics sometimes disagree and do not always perfectly capture the integrity of a model. The statistical metrics for GVDA are certainly less consistent than at other sites in terms of indicating a single best model, albeit Full Rayleigh damping seems to perform the best overall.

## 7. Discussion

### *7.1 Effects of inflated $D_{min}$*

Until subsurface models can more comprehensively capture spatial variability across the large areas and depths impacting site response, thereby allowing for all sources of apparent damping to be explicitly modeled, it appears that empirical damping adjustments will remain a valuable tool for improving the accuracy of GRAs. Small-strain damping adjustments have been shown to serve as a practical means to account for unmodeled dissipation mechanisms and to enhance site response predictions. The results from this study show that applying inflated $D_{min}$ through a damping multiplier significantly enhances agreement between TTFs and ETFs at $f_0$. Importantly, this study establishes a potential relationship between the damping multiplier, $\mathcal{m}$, and the site's velocity contrast (as shown in Figure 9), providing a practical means to estimate the appropriate $\mathcal{m}$ value even in the absence of empirical ground motion recordings at a downhole array. This finding has the potential to be highly impactful, as it offers a straightforward, correlation-based adjustment that can meaningfully improve 1D and 2D GRAs. This relationship between $\mathcal{m}$ value and velocity contrast should be evaluated at additional downhole array sites.

### *7.2 Effects of damping formulation type*

The widespread use of Full Rayleigh damping in numerical GRAs stems from the desire to model frequency-independent damping. Maxwell damping has been proposed as an alternative to Full Rayleigh damping that allows for frequency-independent damping over an even broader frequency range. Results from the present study demonstrate that neither Full Rayleigh nor Maxwell damping produced the best agreement between TTFs and ETFs at three of the four sites. Both formulations, when applied with inflated $D_{min}$ values, consistently overdamped higher frequencies, leading to systematic underprediction of site response. For Maxwell damping, this limitation was further exacerbated by frequency shifts in the modal peaks.

Rayleigh Mass damping provided the closest agreement between TTFs and ETFs at three of the four downhole array sites considered in this study. Rayleigh Mass damping is inherently frequency dependent, applying more damping at low frequencies and less at high frequencies. When combined with $\mathcal{m}$ value, that characteristic proved beneficial because it reduced the excessive damping of higher-mode peaks observed with the other formulations. At GVDA, however, Rayleigh Mass damping did not outperform the other formulations in terms of statistical metrics. This may partly reflect the weaker alignment of TTF and ETF spectral shapes at $f_2$ and $f_3$, regardless of damping formulation, where measures such as the Pearson correlation coefficient are less reliable indicators of accuracy. Since the GVDA pseudo-3D Vs model incorporated fewer and less spatially distributed H/V measurements, a more refined model may be needed to allow more definitive comparisons of damping formulations at this site.

These findings challenge the long-standing assumption that frequency-independent damping is inherently superior for GRAs, and underscore the importance of evaluating damping schemes not only on theoretical grounds, but also on their demonstrated ability to replicate recorded ground motions at downhole array sites. To further evaluate the physical relevance of the damping behavior observed in this study, the frequency-dependent trends obtained using Rayleigh Mass damping were compared with findings from prior studies. Pretell et al. (2023a) reported frequency-dependent small-strain damping trends, showing that increasing $D_{min}$ improves 1D GRA predictions for 1D-like sites, with higher multipliers ($\mathscr{m} > 6$) improving agreement at $f_0$ and lower multipliers being more appropriate at higher frequencies. This behavior is consistent with the approach used for modeling damping in this study, where a $D_{min}$ multiplier is used to inflate damping at low frequencies and Rayleigh Mass damping is used to reduce the inflated damping at higher frequencies. The trends observed in this study were also compared with empirical observations of frequency-dependent damping reported by (Kokusho 2017). As shown in Figure 19, the $\mathscr{m} \times D_{min}$ Rayleigh Mass damping curves used at the four downhole sites closely follow the functional form of frequency-dependent damping proposed by Kokusho (2017), in which equivalent damping decreases with frequency according to $D = 10\% \times f/{f_0}^{-0.8}$. Note that in order to make this comparison, a single representative Rayleigh Mass damping curve needed to be obtained for each downhole array site. This was done by computing a weighted average of the $\mathscr{m} \times D_{min}$ Rayleigh Mass curves across all soil layers. Kokusho (2017) obtained the frequency-dependent damping trend as follows: first, ETFs were computed from ground motions recorded at a large number of KiK-net vertical array sites. Then, equivalent damping ratios were back-calculated by identifying the level of damping required for 1D analytical TTFs to separately match the observed amplification at the $f_0$ peak and at a higher-mode peak in the ETFs. This TTF-to-ETF calibration resulted in empirical damping estimates that varied with frequency. Notably, the analyzed motions included cases with maximum surface accelerations exceeding approximately 0.1–0.2 g, indicating the likelihood of moderate strain-dependent soil nonlinearity. Despite differences in site conditions, strain range, and modeling approaches, the Rayleigh Mass damping curves used herein show strong agreement in both shape and rate of decay with the trend reported by Kokusho (2017), suggesting consistency with in-situ behavior inferred from vertical array observations and supporting the broader applicability of these trends across different site conditions.

Furthermore, several studies focused on nonlinear 1D GRAs (e.g., Assimaki & Kausel, 2002; Huang et al., 2018; Kausel et al., 2002; Kuo et al., 2021; Meite et al., 2020; Yoshida et al., 2002) have also demonstrated that incorporating frequency-dependent damping improves the accuracy of 1D site response predictions. Collectively, these findings support the use of frequency-dependent damping in site response analyses, and the $\mathscr{m} \times D_{min}$ Rayleigh Mass damping formulation presented herein provides an effective means of implementing it.

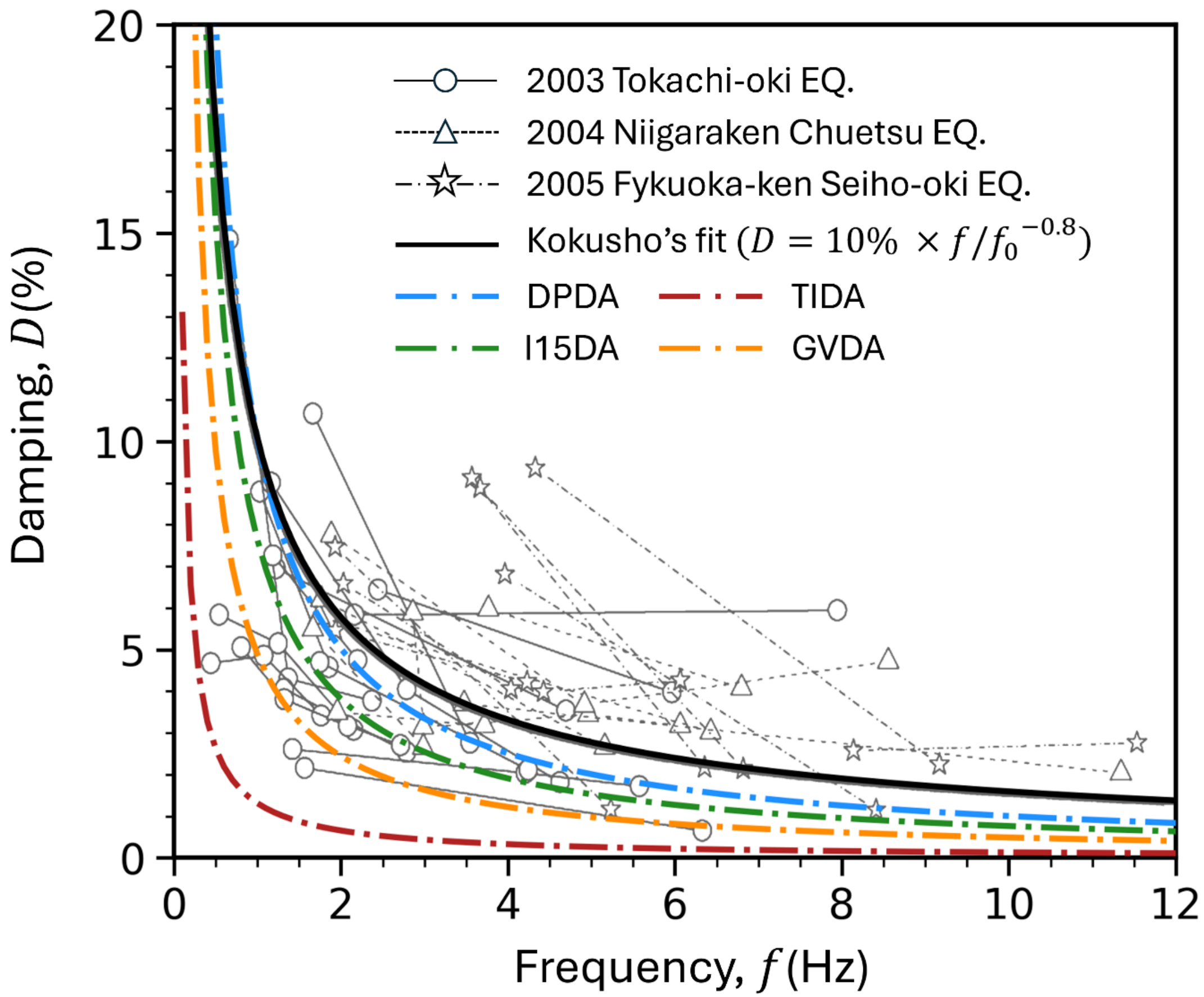

**Figure 19.** Comparison of site-specific ɱ × $D_{min}$ Rayleigh Mass damping curves used in this study with equivalent damping ratios inferred from KiK-net vertical array records in Japan (after Kokusho, 2017). Markers represent damping ratios, as reported by Kokusho, required for analytically computed TTFs to match the amplitudes of observed ETF peaks. Marker pairs connected by straight lines correspond to values from the same site, associated with the $f_0$ peak and a higher-mode peak. The site-specific Rayleigh Mass curves are obtained using a weighted average of layer-specific Rayleigh Mass damping curves at each downhole array site (DPDA, TIDA, I15DA, GVDA).

### *7.3 Frequency shifts induced by Maxwell Damping Formulation*

Observations from the four sites included in this study reveal that Maxwell damping induces noticeable shifts in the modal frequencies. These shifts become increasingly pronounced at higher frequencies. As a result, sites with higher $f_0$, such as GVDA (1.8 Hz) compared to TIDA (0.75 Hz), are more susceptible to these frequency shifts. While the effect may be less severe at sites with lower $f_0$, Maxwell damping consistently introduces peak shifts at higher modes, ultimately impacting the overall accuracy of site response predictions. Moreover, this frequency shift is significantly exacerbated when higher damping levels are used. This is evident in the larger discrepancies observed when ɱ × $D_{min}$ is applied, compared to the use of conventional $D_{min}$ alone. For sites with inherently higher $D_{min}$ values, the magnitude of the shift becomes even more pronounced, highlighting a key limitation of the Maxwell damping formulation. The underlying cause of this behavior is that the Maxwell model not only introduces attenuation, but also makes the real part of the complex modulus frequency dependent. In the Dawson and Cheng formulation (Dawson and Cheng, 2021), each Maxwell branch contributes a term to the complex stiffness whose real part increases with frequency. As a result, the effective stiffness, and therefore the phase velocity, increases across the nominally constant-damping band. Because resonant frequencies scale with wave speed, this increase in phase velocity shifts the transfer-function peaks toward higher frequencies. To the best of our knowledge, no prior studies have directly examined this phenomenon by comparing TTF peak locations to ETF peaks observed at downhole array sites. This study, therefore, presents novel observations of numerical peak shifts induced by Maxwell damping in comparison with recorded ground motions.

### *7.4 Computation Time of Damping Formulations*

In FLAC3D, the size of the critical timestep governs the stability and efficiency of dynamic simulations. The calculation of this timestep is described in the FLAC3D manual (Itasca Consulting Group 2023). When only Rayleigh Mass damping is applied, the timestep remains relatively large and stable, making it computationally efficient. However, once a stiffness-proportional component is added, as in Full Rayleigh damping, the system becomes artificially stiff at higher frequencies. This elevates the maximum eigenfrequency and forces the use of much smaller timesteps to maintain stability. As a result, Full Rayleigh damping is highly sensitive to the damping ratio, with timestep decreasing sharply as $D_{min}$ is inflated. In contrast, Maxwell damping allows timesteps that are comparable to those of Rayleigh Mass damping. The Maxwell damping scheme is implemented in FLAC3D using an implicit algorithm and, therefore, remains unconditionally stable (Dawson and Cheng, 2021). It does not impose additional dynamic restrictions on the numerical timestep. This feature significantly improves computational efficiency, especially in large-scale dynamic analyses. While this makes Maxwell highly efficient from a computational standpoint, its tendency to shift modal peaks significantly reduces its practical value for accurate site response analyses.

The critical timesteps recorded during 2D numerical GRAs for different damping ratios and formulations across the four sites are shown in Figure 20. As expected, distinct differences emerged among the damping schemes. The timestep for Full Rayleigh damping exhibited strong sensitivity to the damping ratio, decreasing sharply as $D_{min}$ was inflated. By contrast, the critical timesteps for Rayleigh Mass and Maxwell damping remained relatively stable across damping ratios at each site. Quantitatively, Rayleigh Mass damping permitted timesteps approximately 2.5–3 times larger than Full Rayleigh damping at $D_{min}$, 5–8 times larger at 2–3 × $D_{min}$ (at TIDA and GVDA), 11 times larger at 8 × $D_{min}$ (at DPDA), and more than 30 times larger at 10 × $D_{min}$ (at I15DA). This substantial increase in allowable timestep translates directly into reduced runtimes, underscoring the superior computational efficiency of Rayleigh Mass damping, particularly at elevated damping levels. In addition to damping formulation, zone size and compressional wave velocity also affect the timestep. However, since these parameters influence all damping formulations equally, their individual effects are not emphasized here. Nevertheless, important site-specific trends are evident. For example, at GVDA, where a very stiff bedrock layer is present, Full Rayleigh damping produced especially small critical timesteps, even when applied with only $D_{min}$. These results highlight that Full Rayleigh damping is the most computationally demanding among the schemes tested. In contrast, Rayleigh Mass damping, particularly when used with inflated $D_{min}$, emerged as the most efficient and practical choice for 2D GRAs.

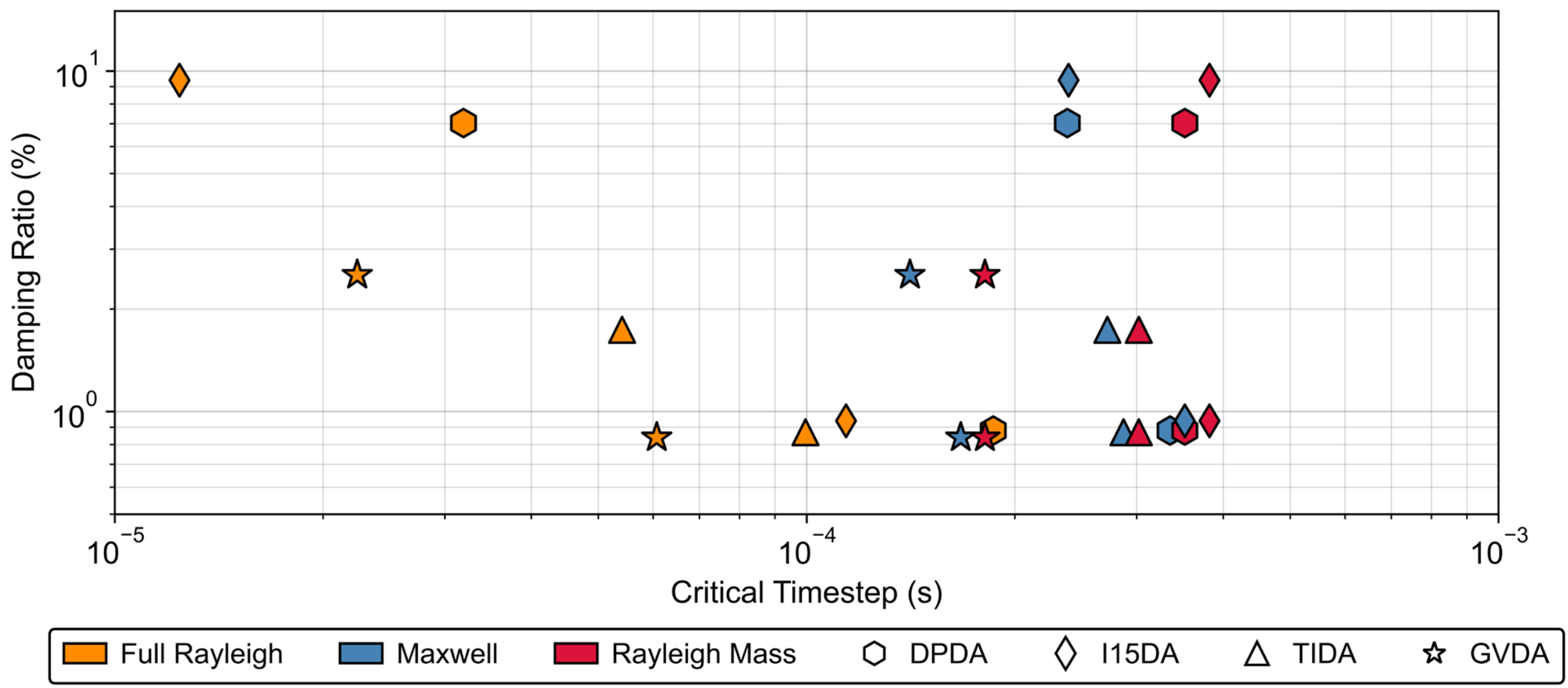

**Figure 20.** Summary of critical timesteps associated with different damping formulations across varying damping ratios.

## 8. Conclusions and Recommendations

This study conducted a comprehensive evaluation of 2D GRAs at four well-instrumented downhole array sites (DPDA, I15DA, TIDA, and GVDA) using three numerical damping formulations: Full Rayleigh, Maxwell, and Rayleigh Mass. Each damping formulation was tested with both conventional $D_{min}$ values and inflated values (ɱ × $D_{min}$) meant to account for unmodeled apparent damping caused by wave scattering, diffractions, and mode conversions. While a full 3D GRA would theoretically provide the most accurate site response predictions, the computational demands of testing six damping formulations across four downhole array sites, resulting in a total of 24 simulation cases, made them impractical. As a practical alternative, 2D GRAs were performed on the most heterogeneous cross-section extracted from pseudo-3D Vs models at each site. For each site, a systematic comparison was made between TTFs generated through linear-viscoelastic numerical simulations and ETFs derived from small-strain recorded ground motions, with an objective of evaluating the effectiveness of these numerical damping formulations in site response predictions.

A key finding is the strong negative correlation between ɱ value and the velocity contrast at each site. At DPDA and I15DA, where velocity contrasts are relatively modest, large ɱ values were necessary to match the 2D GRA predictions with the recorded response, suggesting that apparent damping at these locations may stem from mechanisms not captured well by the pseudo-3D Vs model, such as wave scattering from subsurface lenses or layer boundaries that are not represented in the base Vs profile used in the H/V Geostatistical Approach. By contrast, at TIDA and GVDA, steeply dipping layers with strong velocity contrasts produced scattering and reverberation. Since these features were already well modeled in the pseudo-3D Vs model, additional empirical damping adjustments caused minimal improvement in 2D site response predictions. Because ɱ value cannot be determined *a priori* in most forward engineering applications due to a lack of recorded ground motion data, the recommended correlation between ɱ value and velocity contrast provides a practical means to estimate an appropriate damping multiplier. While the dataset includes only four observations and is thus limited in statistical strength, the consistent site response trends observed support the utility of this correlation. Future studies incorporating more sites can improve the robustness of this relationship.

Full Rayleigh Damping and Maxwell Damping are intended to provide frequency-independent damping schemes for numerical modeling. When applied with the conventional $D_{min}$, both schemes consistently overpredicted peak amplitudes at $f_0$ and at higher modes. When paired with $m$ values to inflate $D_{min}$, they improved agreement at the $f_0$ peak but severely underpredicted the amplitude of higher-mode peaks due to consistent overdamping at high frequencies. In contrast, Rayleigh Mass Damping, a frequency-dependent formulation, provided the closest overall agreement with the observed response at three of the four downhole array sites, especially when combined with inflated $D_{min}$. By applying proportionally less damping at higher frequencies, it effectively avoided the overdamping issues shown by the frequency-independent formulations, allowing the higher-mode amplitudes to be preserved while still accurately capturing the amplitude of the $f_0$ peak. In addition to providing the closest match to recorded response, Rayleigh Mass damping also offered far greater computational efficiency than Full Rayleigh damping, making it an attractive option for large-scale analyses.

These findings have important implications. They challenge the long-standing assumption that frequency-independent damping is inherently preferable for numerical GRAs, showing that frequency-dependent schemes can provide more realistic results when comparisons are made with actual recorded ground motions at downhole arrays sites. More broadly, this study demonstrates that combining pseudo-3D Vs models with calibrated $m$ value and careful selection of damping formulations offers a practical and effective framework for improving multi-dimensional GRAs. Looking forward, these results also provide guidance for future extensions into fully nonlinear 2D and 3D analyses, where a similar balance between numerical efficiency, damping representation, and physical realism will be essential.